\documentclass [12pt]{article}

\def\e{\hbox{E}}

\def\var{\hbox{Var}}

\def\min{\hbox{min}}

\def\hw{\hat{w}}

\def\hth{\hat{\theta}}

\def\d{\partial}

\usepackage{epsfig}
\usepackage{longtable}
\usepackage{natbib}
\usepackage{graphicx}
\begin{document}
\begin{center}
\Large   Testing for Homogeneity in Meta-Analysis\\
I.  The One Parameter Case: Standardized Mean Difference\\

\vspace{1.4cm}
Elena Kulinskaya,  Michael B. Dollinger and Kirsten Bj\o rkest\o l \\
\vspace{1.4cm}1 August 2009
\end{center}
\normalsize \begin{center}
Abstract
\end{center}
\emph{Meta-analysis seeks to combine the results of several experiments in order to improve the accuracy of decisions.  It is common to use a test for homogeneity to determine if the results of the several experiments are sufficiently similar to warrant their combination into an overall result.  Cochran's $Q$ statistic is frequently used for this homogeneity test.  It is often assumed that $Q$ follows a chi-square distribution under the null hypothesis of homogeneity, but it has long been known that this asymptotic distribution for $Q$ is not accurate for moderate sample sizes.  Here we present formulas for the mean and variance of $Q$ under the null hypothesis which represent $O(1/n)$ corrections to the corresponding chi-square moments in the one parameter case.  The formulas are fairly complicated, and so we provide a program (available at \textbf{ http://www.imperial.ac.uk/stathelp/researchprojects/metaanalysis}) for making the necessary calculations.  We apply the results to the standardized mean difference (Cohen's $d$-statistic) and consider  two approximations:  a gamma distribution with estimated shape and scale parameters and the chi-square distribution with fractional degrees of freedom  equal to the estimated mean of $Q$. We recommend  the latter distribution as an approximate distribution for $Q$ to use for testing the null hypothesis. }\\
\vspace{1cm}


\small
\vspace{1 cm}
\noindent
Elena Kulinskaya, Statistical Advisory Service, Imperial College, London, UK.\\
Michael B. Dollinger, Department of Mathematics, Pacific Lutheran University, Tacoma, WA,  USA. \\
Kirsten Bj{\o}rkest{\o}l, Faculty of Technology and Science, University of Agder, Kristiansand, Norway\\

\normalsize
\pagebreak

\section{Introduction} \label{sec:intro}
In the meta-analysis of several studies, it
is usual to conduct a ``homogeneity test'' to determine if the
effects measured by the studies are sufficiently similar to
warrant their combination into one grand summary effect using the fixed effect model, \cite{Normand}.  The most
commonly used test statistic is Cochran's $Q$ \citep{coch-1937}.  It is defined as follows.  Suppose that there
are $I$ studies (or experiments) each of whose result is given by an
estimator $\hat{\theta}_i$ of a population effect $\theta_i$.
Suppose that the variance of $\hat{\theta}_i$ is given by $v_i$
which can be estimated in turn by $\hat{v}_i$.  The summary effects may be combined into a grand summary effect using a
weighted average  $\hat{\theta}_w=\sum_i
\hat{w}_i\hat{\theta}_i/\sum_i \hat{w}_i$ where the weights $w_i$
and their estimators $\hat{w}_i$ are usually taken as inverses of
the variances and their estimators respectively (thus weighting
more accurate studies more heavily).  At this point of
the discussion, the summary effect may be quite general, such as
the sample mean of each study, the difference of means between treatment and control
arms of each study or the standardized difference of means between
treatment and control arms of each study; but in the main body of the paper we will restrict the discussion
to cases in which the estimators of $\theta_i$ and $w_i$ depend on
only the one parameter $\theta_i$.  \\

Cochran's $Q$ statistic, which is used in the homogeneity test, is
defined by $Q=\sum_i\hat{w}_i(\hat{\theta}_i-\hat{\theta}_w)^2$.
When testing the null hypothesis that $\theta_1=\cdots=\theta_I$,
that is the underlying effects measured by all the studies are the
same, it is common to assume that $Q$ has a chi-square
distribution with $I-1$ degrees of freedom.  This distribution
appears to be asymptotically valid (as the sizes $n_i$ of the
studies become large) over a wide choice of summary effects.
There have been many simulation studies of the accuracy of the
chi-square approximation
(see \cite{HO-1985}, \cite{VW-2007a} and the references therein), but except for the case
where the populations are normally distributed with the parameters
estimated by sample means and sample variances, there are few
theoretical results dealing with the question of the distribution
of $Q$ for small or moderate sample sizes.\\

The chi-square distribution is an exact distribution of the $Q$ statistic for normally distributed populations with {\it known} variances resulting in known weights. Randomness of the weights is traditionally ignored in meta-analysis, \cite{B-T-1997},  \cite{J-2006}, \cite{B-J-2008}.
In contrast, Cochran, as early as in his 1937 paper which dealt with the
normally distributed case, recognized the need for a correction to
the chi-square distribution for moderate sample sizes and proposed
such a correction at that time. In 1951, \cite{James-1951} and
\cite{Welch-1951} proposed separate improved corrections to the distribution
of $Q$ (again for the normal  case), corrections which are
equivalent to each other up to order $1/n_i$.  Welch's proposal (more commonly
used and now known as the Welch test) referred $Q$ to a rescaled
$F$-distribution ($cF_{I-1,\nu}$) with $I-1$ and $\nu$ degrees of freedom where
$\nu$ and the rescaling constant $c$ are quantities to be estimated
from the data. In Welch's derivation, the properties of normality including the independence of the estimators of the weights (inverses of sample variances) and of the effects (sample means) was heavily used; these properties are not generally valid in many situations in which the $Q$ statistic is commonly used. Improved approximations to power of the Welch test in the normal case are given in \cite{K-S-G-2003}.  The paper \cite{K-D-K-G-2004} extended the Welch test in the normal case to contrasts (such as the difference of treatment and control means), and a Welch type $Q$ test for robust estimators of effects and their variances was introduced in \cite{ K-D-2007}.\\

In a series of papers, we plan to investigate corrections to the
distribution of $Q$ in situations in which the estimators of the
effects and of the weights are not statistically independent.  As
far as we know, there have been no theoretical results before now
on this subject.  We expect that the results will provide more
accurate homogeneity tests when the sample sizes are small or
moderate.  In this paper (the first of the planned series), we
investigate the situation in which the effect and weight
estimators depend on a single parameter.  We will apply our
general theory to an important special case: the standardized
mean difference (also known as Cohen's $d$, \cite{Cohen-1988}).
Definitions appear in Section~\ref{sec:SMD}. \\

This paper is organized as follows.  In Section \ref{sec:theory},
we present the general theory.  In Section \ref{sec:SMD} we apply
the general theory to the standardized mean difference. Section
\ref{sec:Example} contains two real meta-analytic examples which have used the standardized mean difference to measure the effects.  Section
\ref{sec:SMDsims} contains the results of a large number of
simulations which show the quality and the limitations of the new
approximations for the homogeneity test based on $Q$ when the
effects are measured as standardized mean differences.  In the final section we summarize the more important conclusions, make some recommendations and indicate areas of future work.  Some of the more complicated formulas have been relegated to the Appendix.

\section{The general theory} \label{sec:theory}
Welch's 1951 correction to the distribution of $Q$ was based on
expansions to approximate the mean and variance of $Q$.  He
then used these moments to define an approximate distribution for
$Q$.  We follow this same general idea, but there are several
important differences.  Welch made the assumption that the underlying distributions were normally distributed and that the weights were inverses of the study variances, estimated by the sample variances.  To permit as wide applicability as possible, we do not assume normality and allow the weights to be different from the inverses of the variances.  Also we do not make the assumption that the estimators of the weights are statistically independent of the estimators for the effects.   A third difference between our approach and that of Welch is that he based  his approximations on an
asymptotic expansion of the moment generating function of $Q$.  We
instead use the delta method, which is based on Taylor expansions
of $Q$ and $Q^2$ about the  mean of the effect size.

\subsection{Notation and assumptions} \label{subsec:notation}
There are $I$ studies with corresponding effect $\theta_i$.  The null hypothesis for the homogeneity test will be equality of the effects, i.e., $\theta_1=\cdots=\theta_I$; we will denote the common effect under the null hypothesis by $\theta$.  The effects are estimated by random variables $\hth_i$.  The theoretical weights are $w_i$ and they are estimated by $\hw_i$. In most applications, we will have $w_i=1/\var[\hth_i]$, but in this section we merely assume that the weight estimators are some function $f_i$ of the effect estimator $\hth_i$; that is, $\hw_i=f_i(\hth_i)$ where the functions $f_i$ will generally depend on additional constants such as the sample size.  The theoretical weights under the null hypothesis will be $w_i=f_i(\theta)$.  The assumption that the weights are dependent only on the corresponding effects is an important limitation of the results of this paper.  In our next paper in this series, we plan to investigate the situation in which the weights depend on more than one random variable.

We need to make some fairly standard assumptions about the orders (relative to the sample sizes) of the central moments $\e[(\hat{\theta}_i-\theta_i)^r]$ of the estimators $\hth_i$ and also the orders of the weights and their derivatives.  Let $n_i$ represent the sample size of the $i$th study.  In the event that the studies have two arms (as in the application in Section~\ref{sec:SMD}), let $n_i$ be the minimum sample size of the two arms.  We will also use the notation $n=\min\{n_i\}$ and sometimes express approximations in terms of orders of $n$.

 To simplify notation, define $\Theta_i=(\hth_i-\theta_i)$.  We assume first that $\e[\Theta_i]=O(1/n_i^2)$. This condition will certainly be satisfied if the estimator $\hth_i$ is unbiased. In regular parametric problems, it is easy to remove the first-order term from the asymptotic bias of maximum likelihood estimates (see \cite{Firth}).  We will need higher moments up to and including the sixth central moment.  For these moments, we assume the following orders which generally follow from $\sqrt{n_i}$ asymptotic normality: $\e[\Theta_i^2]=O(1/n_i)$, $\e[\Theta_i^3]=O(1/n_i^2)$, $\e[\Theta_i^4]=O(1/n_i^2)$, $\e[\Theta_i^5]=O(1/n_i^3)$ and $\e[\Theta_i^6]=O(1/n_i^3)$.  We further assume that the weight estimators $\hw_i$ and their first two derivatives with respect to $\theta_i$ will be $O(n_i)$, as will be the case whenever the weights are inverses of the variances.

\subsection{Expansions for $\e[Q]$ and $\e[Q^2]$} \label{subsec:equations}
In this section we present expressions for $\e[Q]$ and $\e[Q^2]$ using Taylor expansions and then taking expectations of these expansions.  The Taylor expansions are centered about the the null hypothesis $\theta_1=\cdots=\theta_I=\theta$, and thus all derivatives in this section are to be evaluated at this null hypothesis.  In our expansions we have kept all terms to order $O(1/n)$.  We begin with the first moment of $Q$.

\begin{eqnarray}
\e[Q] &=&\frac{1}{2}\sum_i \frac{\d^2Q}{\d\theta_i^2} \e[\Theta_i^2]  + \frac{1}{6}\sum_i \frac{\d^3Q}{\d\theta_i^3}\e[\Theta_i^3] \\
 &+& \frac{1}{24}\sum_i\frac{\d^4Q}{\d\theta_i^4}\e[\Theta_i^4] + \frac{1}{8}\sum_{i\neq j}\sum\frac{\d^4[Q]}{\d\theta_i^2\d\theta_j^2}\;
\e[\Theta_i^2]\e[\Theta_j^2] +O\left(\frac{1}{n^2}\right) \nonumber
\end{eqnarray}

We next substitute expressions for the indicated derivatives into this formula and expand the double sum into combinations of single sums to obtain the following result.  To simplify the expression, we use the notation $W=\sum_iw_i$ and $U_i=1-w_i/W$.  The formula is expressed in terms of parameter values; estimates of these parameter values will be needed when the formula is applied to data.

\begin{eqnarray}
\e[Q] &= &\sum_i [w_i U_i] \e[\Theta_i^2]  + \sum_i \left[U_i^2 \frac{d f_i}{d \hth_i}\right]\e[\Theta_i^3]
 + \sum_i\left[-\frac{U_i^2}{W}\left(\frac{d f_i}{d \hth_i}\right)^2 +
 \frac{U_i^2}{2}\frac{d ^2 f_i}{d \hth_i^2}\right]\e[\Theta_i^4] \nonumber\\
 &-&\frac{1}{W}\left(\sum_i U_i\frac{d f_i}{d \hth_i}\e[\Theta_i^2]\right)^2 -  \frac{1}{W^3}\left(\sum_iw_i\frac{d f_i}{d \hth_i}\e[\Theta_i^2]\right)^2 \label{eq:EQ1}\\
&-&\frac{1}{W^3}\left(\sum_iw_i\e[\Theta_i^2]\right) \left(\sum_i\left\{\left[\frac{d f_i}{d \hth_i}\right]^2
-\frac{1}{2W}\frac{d^2f_i}{d \hth_i^2}\right\}\e[\Theta_i^2]\right) \nonumber \\
&+&\frac{1}{W}\sum_i\left(1-\frac{2w_i}{W}+
\frac{3w_i^2}{W^2}\right)\left[\frac{d f_i}{d \hth_i}\right]^2(\e[\Theta_i^2])^2
-\frac{1}{2W^3}\sum_iw_i^2\frac{d^2f_i}{d \hth_i^2}(\e[\Theta_i^2])^2 +O\left(\frac{1}{n^2}\right) \nonumber
\end{eqnarray}

The expansion for second moment $\e[Q^2]$ up to order $O(1/n)$ requires terms of 4th, 5th and 6th degree.  The expansion is given by
\begin{eqnarray}
\e[Q^2] &=& \frac{1}{24}\sum_i \frac{\d^4[Q^2]}{\d\theta_i^4}\; \e[\Theta_i^4]
+ \frac{1}{8}\sum_{i\neq j}\sum\frac{\d^4[Q^2]}{\d\theta_i^2\d\theta_j^2}\;
\e[\Theta_i^2]\e[\Theta_j^2]
+ \frac{1}{120}\sum_i \frac{\d^5[Q^2]}{\d\theta_i^5}\;\e[\Theta_i^5] \nonumber \\ &+&
\frac{1}{12}\sum_{i \neq j}\sum \frac{\d^5[Q^2]}{\d\theta_i^3\d\theta_j^2} \;
\e[\Theta_i^3]\e[\Theta_j^2]
+\frac{1}{720}\sum_i \frac{\d^6[Q^2]}{\d\theta_i^6}\;\e[\Theta_i^6] \label{eq:EQ2}\\
&+&\frac{1}{48} \sum_{i \neq j}\sum  \frac{\d^6[Q^2]}{\d\theta_i^4\d\theta_j^2}  \;
\e[\Theta_i^4]\e[\Theta_j^2]
 +\frac{1}{48}\sum_{i \neq j}\sum_{\neq
k}\sum\frac{\d^6[Q^2]}{\d\theta_i^2\d\theta_j^2\d\theta_k^2} \;
\e[\Theta_i^2]\e[\Theta_j^2]\e[\Theta_k^2] +O\left(\frac{1}{n^2}\right)\nonumber
\end{eqnarray}

The derivatives of $Q^2$ needed for Equation ~\ref{eq:EQ2} are fairly complicated and appear in the Appendix.

\subsection{Applying the formulas}\label{sec:applyformulas}

The formulas in Equations~\ref{eq:EQ1} and \ref{eq:EQ2} are fairly general since they are not based on any normality assumptions, and will be applicable to any situation in which there is only one parameter and in which the weights and central moments meet the order conditions described in Section~\ref{subsec:notation}.\\

To use the formulas for a specific application, the user will need to supply expressions for the weights (that is, the functions $f_i$) and their first and second derivatives and also expressions for the central moments $\e[\Theta_i^r]$ for $r=1,\ldots,6$.  We provide an illustration in the next section where we apply the theory to the important special case of the standardized mean difference.  Because of the complexity of the formulas, we have provided a computer program in $R$ which can be used for the necessary calculations for applying the $Q$ test to the standardized mean difference.  This program can be downloaded from the website  \textbf{ http://www.imperial.ac.uk/stathelp/researchprojects/metaanalysis}.\\

The weights and their derivatives which appear in the formulas need to be estimated under the null hypothesis and will be different from the weights which are used for calculating a specific value of $Q$ from the data.  Specifically, weights $\hw_i=f_i(\hth_i)$ are first calculated. These weights are used to estimate the combined effect $\hth_w=\sum_i
\hat{w}_i\hat{\theta}_i/\sum_i \hat{w}_i$ and to calculate the value of the $Q$ statistic $\sum_i\hat{w}_i(\hat{\theta}_i-\hat{\theta}_w)^2$.  However, the weights which appear in Equations~\ref{eq:EQ1} and \ref{eq:EQ2} need to be recalculated using the same combined effect $\hth_w$ as the effect for each of the studies.  That is, these `null' weights are estimated by $f_i(\hth_w)$ and the derivatives will be estimated by $\frac{\d f_i}{\d \theta_i}(\hth_w)$ and $\frac{\d^2f_i}{\d \theta_i^2}(\hth_w)$. \\

Improved approximations to the mean and variance of $Q$ under the null hypothesis are, of course, not sufficient to conduct a test of the null hypothesis.  A distribution for $Q$ is needed for this purpose.  Ideally, simulations should be used for each separate application type to select a family of distributions which fits the distribution of $Q$.  However, we have found in our simulations, which cover a number of situations (including both the one parameter case discussed here as well as in cases involving multiple parameters), that the gamma family of distributions fits the null distribution of $Q$ quite closely.  Importantly, this family includes the chi square family as a special case.  In particular, we have found that the gamma family of distributions fits the distribution of $Q$ very well in the case of the effects are measured by the standardized mean difference. Another contender is the chi-square distribution with fractional degrees of freedom equal to the mean of $Q$ (see Section~\ref{sec:gammaSMD} below).

\subsection{Inverse variance weights and the chi-square distribution}

It is usual to choose weights to be inverse variances, i.e., $w_i=1/\e[\Theta_i^2]$.  We make this assumption in the remainder of this section.  The expressions for the moments given in Equations~\ref{eq:EQ1} and \ref{eq:EQ2} simplify somewhat under this inverse variance assumption.  In Equation~\ref{eq:EQ1}, only the first (or quadratic) term is $O(1)$.  The remaining terms are all $O(1/n)$.  With inverse variance weights, this first term simplifies to $\sum_i(1-w_i/W)=I-1$.  Notice that this quantity is the first moment of the chi-square distribution with $I-1$ degrees of freedom.  Thus Equation~\ref{eq:EQ1} provides an order $O(1/n)$ correction to the chi-square first moment.

In Equation \ref{eq:EQ2} for the second moment of $Q$, the lowest degree terms
are the first two terms (those of fourth degree), and these are the only two terms of order $O(1)$.  The remaining terms are all of order $O(1/n)$.  Using Equations~\ref{eq:d^4(Q^2)} and \ref{eq:d^4(Q^2ij)} (in the Appendix) for the fourth derivatives of $Q^2$, these two terms become
\begin{equation}
\sum_iw_i^2U_i^2\e[\Theta_i^4] + \sum_{i\neq j}\sum(U_iU_j+\frac{2w_iw_j}{W^2}).
\label{eq:EQ^2O(1)}\end{equation}
The kurtosis $\gamma_2$ of a random variable with fourth central moment $\mu_4$ and variance $\sigma^2$ is commonly defined by $\gamma_2=\mu_4/\sigma^4-3$; this definition is arranged so that normally distributed random variables have kurtosis of zero.  Using this definition, we will denote the kurtosis of $\hth_i$ (the estimator of the $i$th effect) by $\gamma_{2,i}$.  Then Equation~\ref{eq:EQ^2O(1)} can be algebraically rearranged to
\begin{equation}
I^2-1+ \sum_i\gamma_{2,i}U_i^2.
\end{equation}
Since kurtosis is typically of order $O(1/n)$, we see that the second moment of the null distribution of $Q$ agrees with the second moment of the chi-square distribution with $I-1$ degrees of freedom (which is $I^2-1$) up to order $O(1/n)$.

Thus when inverse variance weights are used, both the first and second moments of the null distribution of $Q$ agree with those of the chi-square distribution up to order $O(1)$ and Equations~\ref{eq:EQ1} and \ref{eq:EQ2} provide order $O(1/n)$ corrections. \\

When discussing the distribution of $Q$, some authors make the simplifying assumption that the weights are constants rather than random variables.  See, for example \cite{B-T-1997}, \cite{J-2006}, \cite{B-J-2008}.  When this assumption of constant weights holds, the derivatives of the weights become zero and all terms of our approximate formula for $\e[Q]$ vanish except for the first (or chi-square) term.  Similarly, all terms for $\e[Q^2]$ vanish except for the first two terms.  Accordingly, under the assumption that the weights are known constants, the commonly used chi-square approximation for $Q$ has mean which is accurate to order $O(1/n)$.  But the second moment is accurate to this order only when the estimators of the effects have kurtosis of order less than $1/n$.  However, since in reality the weights are random, both the mean and variance of $Q$ need the corrections given by our formulas in order to be accurate to order $O(1/n)$.  Thus use of our formulas should yield improved accuracy in the $Q$ test when $n$ is not too large.

\section{The $Q$ test for the standardized mean difference} \label{sec:SMD} In this section, we apply the theoretical results of the previous section to the standardized mean difference (also known as Cohen's $d$-statistic).  We begin with notation and a brief review of the necessary background.  See, for example, \cite{HO-1985} for details.

\subsection{Notation and weight functions}\label{sec:SMDnotation}
We assume that each of $I$ studies consists of two arms of sizes
$n_{Ti}$ and $n_{Ci}$ having normally distributed data with
means $\mu_{Ti}$ and $\mu_{Ci}$ and that the variance $\sigma_i^2$
is the same in each arm.  (The subscripts $T$ and $C$ may be thought of as treatment and control.)  Then the effect measured by the standardized mean difference in the $i$th study is given by
\begin{equation}
\delta_i=(\mu_{Ti}-\mu_{Ci})/\sigma_i.
\end{equation}
 A natural, but biased, estimator of $\delta$ is
\begin{equation}
\hat{\delta}_i=(\bar{X}_{Ti}-\bar{X}_{Ci})/s_{pi}
\end{equation}
where $s^2_{pi}$ is the usual pooled variance estimator.  Instead of using $\hat{\delta}_i$, we follow the usual practice to correct for the bias by using the unbiased estimator of $\delta$ defined by
\begin{equation}
\hat{g_i}=J_i\hat{\delta}_i=J_i(\bar{X}_{Ti}-\bar{X}_{Ci})/s_{pi}
\end{equation}
where
\begin{equation}
J_i=\frac{\Gamma[(N_i-2)/2]}{\sqrt{(N_i-2)/2}\;\;\Gamma[(N_i-3)/2]}
\end{equation}
is a constant depending only on the total sample size
$N_i=n_{Ti}+n_{Ci}$.    Define $q_i=n_{Ci}/N_i$  to be the proportion of
the total sample size in the control arm of the $i$th study. It is
known that (see  \cite[p. 104--5]{HO-1985})
\begin{equation}\label{eq:var(g)}
\var[\hat{g_i}] = \frac{(N_i-2)J_i^2}{(N_i-4)N_iq_i(1-q_i)}+
\left(\frac{(N_i-2)J_i^2}{N_i-4}-1\right)\delta_i^2:=A_i+B_i\delta_i^2,
\end{equation}
where the constants $A_i$ and $B_i$ depend only on the sample
sizes.  Replacing $\delta_i$ by its unbiased estimator $\hat{g}_i$ in this variance formula, we obtain an estimator of the variance of $\hat{g}_i$ which is given by
\begin{equation}
\hat{\var}[\hat{g}_i]=A_i+B_i\hat{g}_i^2.
\end{equation}
Then the functions $f_i$ giving the estimated inverse variance weights in the $Q$ statistic are given by
\begin{equation}
\hat{w_i} = f_i(\hat{g}_i) =\left[A_i+B_i\hat{g}_i^2\right]^{-1}.
\label{eq:SMDweights}
\end{equation}
The first and second derivatives of $\hat{w}_i$ are given by
\begin{eqnarray}
\frac{df_i}{d\hat{g}_i}&=& -2B_i\hat{g}_i\hat{w}_i^2 \\
\frac{d^2f_i}{d\hat{g}_i^2}&=&-2B_i\hat{w}_i^2+8B_i^2\hat{g}_i^2\hat{w}_i^3.
\end{eqnarray}\\

One issue that has arisen in meta-analysis involving the standardized mean difference is how best to estimate the combined effect $\delta$.  Estimators of $\delta$ appear in two places in the $Q$ test: in the definition of $Q$; and in the application of Equations~\ref{eq:EQ1} and \ref{eq:EQ2} where an estimated  value of $\delta$ under the null hypothesis is used.  It is known (see \cite{Yuan}) that the natural weighted sum estimator $\hat{g}_w=\sum \hat{w}_i\hat{g}_i/\sum \hat{w}_i$ is slightly biased.  An alternative choice is to use the estimator $\hat{g}_A =\sum A_i\hat{g}_i/\sum A_i$; since the weights $A_i$ are not random, the estimator $\hat{g}_A$ is unbiased.  We explored both choices in our simulations of the $Q$ test and found that the difference between these two choices is barely noticeable and not of practical importance.  We use the estimator $\hat{g}_w$ in the examples of Section~\ref{sec:Example}.

\subsection{The moments of $\hat{g}$}
In this section we suppress the subscript $i$ on all variables pertaining to the $i$th study.  The two main ingredients needed for applying the formulas for $\e[Q]$ and $\e[Q^2]$ are first the weight functions and their derivatives (given in the previous section) and second the central moments
$\e[(\hat{g}-\delta)^r]$ for $r=1,\ldots,6$. We provide these central moments in this section.  For these moments to exist, we assume that $N>8$. (We note that for the usual chi-square approximation to hold, $N>4$ is required just for the variance of $g$ to exist.) It is known that
(\cite[p. 79]{HO-1985}) $\sqrt{(Nq(1-q))}\;\hat{\delta}$ has a non-central
$t$-distribution with $N-2$ degrees of freedom and non-centrality
parameter equal to  $\sqrt{(Nq(1-q))}\;\delta$.  To simplify
notation, write  $\gamma= \sqrt{(Nq(1-q))}\;\delta$ for the non-centrality parameter.

Denote a random variable with a non-central $t$-distribution with
$N-2$ degrees of freedom and non-centrality
parameter $\gamma$ by $t_{N-2}(\gamma)$.  Then from \cite[p. 512]{J-K-B-1995}, the moments of $t_{N-2}(\gamma)$ about zero
are given by
\begin{equation}
\e[t_{N-2}^r(\gamma)]=\left(\frac{N-2}{2}\right)^{r/2}
\frac{\Gamma[\frac{N-2-r}{2}]}{\Gamma[\frac{N-2}{2}]} \sum_{j=0}^{\lfloor r/2 \rfloor}
{r \choose 2j}\frac{(2j)!}{2^jj!}\gamma^{r-2j}.
\end{equation}
The first moment of $t_{N-2}(\gamma)$ will be denoted
 by $\mu_t$ and is given by
\begin{equation}
\mu_t=\left(\frac{N-2}{2}\right)^{1/2}
\frac{\Gamma[\frac{N-3}{2}]}{\Gamma[\frac{N-2}{2}]}\;\gamma
\end{equation}
Then the central moments of $t_{N-2}(\gamma)$ are given by
\begin{equation}
\e[\left(t_{N-2}(\gamma)-\mu_t\right)^r]= \sum_{k=0}^r (-1)^k {r \choose k} \mu_t^k
\e[t_{N-2}^{r-k}(\gamma)]
\end{equation}

Since $\frac{\sqrt{(Nq(1-q))}}{J}\;\hat{g}$ has the distribution
$t_{N-2}(\gamma)$, we then have the desired central moments needed
for the formula for $Q$.  These are
\begin{equation}\label{eq:momg}
\e[(\hat{g}-\delta)^r]=\left(\frac{J}{\sqrt{(Nq(1-q))}}\right)^r
\e[\left(t_{N-2}(\gamma)-\mu_t\right)^r]
\end{equation}

\subsection{Verifying the order conditions}
One further step in applying the formulas for $\e[Q]$ and $\e[Q^2]$ is to check the order conditions which are set out in Section~\ref{subsec:notation}.  Recall that we use the notation $N_i$ to represent the sum of the sizes of the two arms of the $i$th study and that we use the notation $n_i$ to be the minimum of the two sizes, with $n=\min\{n_i\}$.  It is evident from the definition in Equation~\ref{eq:var(g)} that $A_i=O(1/n)$.  Also $B_i$ (as defined in Equation~\ref{eq:var(g)}) is $O(1/n)$; see \cite{HO-1985} for this fact.  Thus $\hat{w_i} = f_i(\hat{g}_i)$ and its derivatives are $O(n)$.  Further, since $\hat{g}_i$  is unbiased, the order condition for the first central moment of $\hat{g}_i$ is trivially satisfied.

In the remainder of this paragraph, we again suppress the subscript $i$ on all variables pertaining to the $i$th study in order to simplify notation.  Let $X$ denote a normally distributed random variable with mean $\gamma$ and variance 1, i.e., $X\sim \textbf{N}(\gamma, 1)$.  Then the $k$th moments of the  noncentral $t_{N-2}(\gamma)$ distribution are related to the moments of $X$ by
\begin{equation}
\mu_k(t_{N-2}(\gamma))=\mu_k(X)\frac{\Gamma[(N-2-k)/2](N-2)^{k/2}}{2^{k/2}\Gamma[(N-2)/2]} \end{equation}
where $\mu_k$ denotes the $k$th moment (see \cite{Bain}). From Stirling's formula, $\mu_k(t_{N-2}(\gamma))=\mu_k(X)(1+O(n^{-1}))$.
Therefore, from equation (\ref{eq:momg}), the central moments of $\hat{g}$ are in the limit (up to an $O(1)$ multiplier $J^r$) the central moments of the $\textbf{N}(\delta,(Nq(1-q))^{-1})$ distribution, so the order conditions are satisfied.

\subsection{The gamma distribution}\label{sec:gammaSMD}
From our many simulations, it has become apparent that the gamma family with probability density functions $$f(t)=\frac{1}{\Gamma(\alpha)\beta^{\alpha}}t^{\alpha-1}e^{-t/\beta}$$ is a very good fit to the distribution of $Q$ under the null hypothesis of equal standardized mean differences.  For a random variable $T$ with a gamma distribution, the shape parameter $\alpha$ is given by $\alpha=(\e[T])^2/\var[T]$ and the scale parameter $\beta$ is given by $\var[T]/\e[T]$.  The chi-square distribution with $\nu$ degrees of freedom is a member of the gamma family with $\alpha=\nu/2$ and $\beta=2$.

To verify the fit of the gamma family to the null distribution of $Q$, we simulated a number of empirical distributions of $Q$ and used the statistics package \emph{Statgraphics Centurion XV} (from Statpoint, Inc.) to compare the fit of these empirical distributions with a variety of distribution families.  The gamma family always was the best, typically with a Kolmogorov-Smirnov (K-S) distance of only 0.002, which indicates a remarkably good fit.  The second best fitting family was the chi-square family with fractional degrees of freedom which typically had a K-S distance of four times that of the best fitting gamma distribution.

\section{Examples}\label{sec:Example}
In this section, we present two examples to illustrate the application of the theory of Sections \ref{sec:theory} and \ref{sec:SMD} to real data.  Our program, available at \\ \textbf{http://www.imperial.ac.uk/stathelp/researchprojects/metaanalysis } can be used to perform the calculations for these examples.

\subsection{Meta-analysis of the use of a placebo for pain relief}
As a first example, consider the meta-analysis by \cite{H-G-2004} of 17 randomized clinical trials comparing the use of a placebo for pain against no treatment at all.  Summary data from the meta-analysis is found in Table~\ref{tab:H-Gdata}.
\begin{table}[h]
 \begin{footnotesize}
   \begin{center}
   \begin{tabular}{|l||r|r|r|r|r|r|r|r|}
     \hline
    Study&$n_T$&$\bar{X}_T$&$s_T$&$n_C$&$\bar{X}_C$&$s_C$&$\hat{g}$&$w$ \% \\
    \hline
   Reading 1982&  18 &1.60&  1.30& 20 & 2.30 &2.00&--0.402&4.3\\
 Conn 1986&  13  &28.20 &18.40 &14&44.40 &15.70&--0.921&2.8\\
 Hashish 1986&  25 &16.00&  11.70 & 50 & 30.00 &18.90&--0.821&7.2 \\
 Hashish 1988 &   25& 42.00& 25.00 &    25 &60.00 &23.00&--0.738&5.4 \\
  Hargreaves 1989 &   25 &4.50 & 2.50 &    25 &4.90 &2.40&--0.161&5.8 \\
   Blanchard 1990b&  18    &11.90 &23.90  & 24    &20.70 &34.80& --0.282&4.7 \\
   Blanchard 1990a & 13   & 8.30& 13.60 & 11    & 22.50&25.10&--0.697& 2.5\\
   Sprott 1993 & 10   &7.90 & 3.00 & 10    &7.40 &3.00&0.160& 2.3\\
   Forster 1994 & 15   &3.20 &2.80  & 15    & 4.60&2.20&--0.541& 3.3\\
    Parker 1995& 49   &4.00 & 1.90 & 45    &3.80 &2.20&0.097& 10.9\\
    Rowbotham 1996& 35   &--4.40 & 8.70 & 35    &1.90 &8.70&--0.716&7.6 \\
    Wang 1997& 25   &10.70 & 7.30 & 26    &13.40 &5.80&--0.404& 5.8\\
    Robinson 2001& 13   &3.85 & 3.48 & 10    &4.25 &3.74&--0.107& 2.6\\
    Cupal 2001& 10   &2.70 &0.95  & 10    &2.70 &1.34&0.000& 2.3\\
    Rawling 2001& 89   &5.30 & 4.72 & 96    &5.60 &4.90&--0.062& 21.6\\
    Kotani 2001& 23   &15.00 &4.50  & 24    &18.00 &6.00&--0.554& 5.2\\
    Lin 2002& 25   &30.20 & 14.40 & 25    & 38.10&16.00&--0.511& 5.6\\
\hline
\end{tabular}
\end{center}
\caption{\label{tab:H-Gdata} \emph{Data on placebo interventions for pain, \cite{H-G-2004}.  The data are on clinician-rated pain scales. The subscripts $T$ and $C$ refer to the treatment and control arms of the studies. The column headed $\hat{g}$ contains the estimated standardized mean differences between the two arms of each study and the column headed $w$ are the weights (as percentages) used in computing the $Q$ statistic.}}
\end{footnotesize}
\end{table}

Because different studies used different measurement scales for evaluating pain, the standardized mean difference is used in the meta-analysis in order to place each of the effects on a scale free basis.  The effect from each study appears in the table in the column headed $\hat{g}$.  The weights (from Equation~\ref{eq:SMDweights}) which appear in the last column of the table are given as percentages for ease of comparison, but the actual weights are needed for computation of the $Q$ statistic.  The actual weights can be computed using the weight total which is $W= 212.91$.  The weighted average of the effects is $\hat{g}_w =-0.338$.  The value of Cochran's $Q$ statistic is 22.07. Using the standard chi-square approximation with 16 degrees of freedom provides the p-value of 0.141 for the test for homogeneity.

To use the results from Sections \ref{sec:theory} and \ref{sec:SMD}, first the weights need to be recalculated to reflect the null hypothesis of equal standardized mean differences.  We take this null value (as found above) to be $\hat{g}_w =-0.338$ for each of the 17 studies and recalculate the weights using Equation~\ref{eq:SMDweights}.  Then the estimated first and second moments of the null distribution of $Q$ can be calculated from Equations~\ref{eq:EQ1} and \ref{eq:EQ2} and the Appendix yielding the values $\e[Q] =15.19$ and $\e[Q^2]=257.57$ respectively.  Thus the estimated parameters of the approximating gamma distribution are $\alpha = 8.96$ (shape parameter) and $\beta=1.70$ (scale parameter).  The p-value corresponding to $Q=22.07$ is 0.098. The p-value for a chi-square distribution with $\e(Q)=15.19$ degrees of freedom is 0.112.

To assess the relative accuracy of the three approximations (gamma and chi-square with 16 and with 15.19 degrees of freedom) to the null distribution of $Q$, we conducted a simulation of 100,000 random samples with seventeen studies having the same sizes as those of \cite{H-G-2004}, but with all studies having the null value of the standardized mean difference $\delta = -0.338$.  The comparisons are as follows, where the notation `true' null refers to the simulation of 100,000 samples:

\begin{center}
\begin{tabular}{|c|c|c|c|c|c|}
\hline
& p-value for $Q=22.07$ & $\e[Q]$ & $\e[Q^2]$& $\alpha$&$\beta$\\
\hline
simulation (`true' null) & 0.108 & 15.22 & 260.76 &&\\
chi-square est-df & 0.112 & 15.19 &&&\\
gamma  & 0.098 & 15.19 & 257.57&8.96&1.70\\
chi-square 16 df & 0.141 & 16 & 288&&\\
\hline
\end{tabular}
\end{center}

The p-value produced by the gamma distribution and especially that from the chi-square distribution with fractional degrees of freedom are substantially closer to the `true' p-value as given by the simulations.  Notice that the first and second moments of the `true' null distribution of $Q$ are smaller than the corresponding moments of the chi-square distribution, indicating the need for corrections.  Our formulas produce an excellent approximation of the first moment.  The approximation for the second moment is much better than that given by the chi-square distribution, but it is not nearly as good as the approximation of the first moment.

\subsection{Meta-analysis of light therapy for depression}
For a second example, consider the data from a meta-analysis of five studies to determine the effect of light therapy for non-seasonal depression (bright light vs standard treatment), \cite{TKE-2004}.  See Table \ref{tab:example} for the summary data.

 \begin{table}[ht]
  \begin{footnotesize}
   \begin{center}
   \begin{tabular}
   {|l||r|r|r|r|r|r|r|r|}
     \hline
    Study&$n_T$&$\bar{X}_T$&$s_T$&$n_C$&$\bar{X}_C$&$s_C$&$\hat{g}$& $w (\%)$ \\
    \hline
   Holsboer 1994&  14 &14.50&  5.59& 14 & 8.64 &8.38&0.80& 23.2\\
 Fritzsche 2001b&  10  &15.80 &5.30 &10&16.90 &6.40&--0.18& 17.8\\
 Fritzsche 2001a&  11 &10.01&  8.60 & 9 & 9.50 &3.80&0.07& 17.7 \\
 Prasko 2002 &   11& 17.00& 11.20 &    9 &13.00 &7.90&0.39 & 17.3 \\
Benedetti 2003 &   18 &11.72 & 9.25 &    12 &18.75 &7.78&--0.79 & 24.0\\
\hline
\end{tabular}
\caption{\label{tab:example} \emph{Data from a meta-analysis of light therapy for non-seasonal depression (bright light vs standard treatment), \cite{TKE-2004}.  The data are on clinician-rated mood scales. The subscripts $T$ and $C$ refer to the treatment and control arms of the studies. The column headed $\hat{g}$ contains the standardized mean differences between the two arms of each study and the column headed $w$ are the weights (as percentages) used in computing the $Q$ statistic.}}
\end{center}
\end{footnotesize}
\end{table}

The outcomes of the treatments were measured on a clinician-rated mood scales.  The standardized mean difference statistic was used in the meta-analysis because different mood-scale scores were used in different studies.  The weighted average of the effects is 0.0437.  The total of the weights is 27.1.  The value of Cochran's $Q$ statistic is 8.86, and the standard chi-square approximation with 4 degrees of freedom  provides the p-value of 0.065 for the test for homogeneity.

To use the results from Sections \ref{sec:theory} and \ref{sec:SMD}, first the weights need to be recalculated to reflect the null hypothesis of equal standardized mean differences.  We take this null value (as found above) to be $\hat{g}_w =0.0437$ for each of the 5 studies and recalculate the weights using Equation~\ref{eq:SMDweights}.   Then the formulas yield the following results. The estimated first and second moments of the null distribution of $Q$ are $\e[Q] =3.70$ and $\e[Q^2]=19.37$ respectively.  Thus the estimated parameters of the approximating gamma distribution are $\alpha = 2.41$ (shape parameter) and $\beta=1.54$ (scale parameter).  The p-values corresponding to $Q=22.07$ are 0.037 (gamma approximation) and 0.053 (chi-square with 3.70 degrees of freedom).

To assess the relative accuracy of the gamma and chi-square approximations to the null distribution of $Q$, we conducted a simulation of 100,000 random samples with five studies of the same sizes as that of \cite{TKE-2004}, but with all studies having the null value of the standardized mean difference $\delta = 0.437$.  The comparisons are as follows where the notation `true' null refers to the simulation of 100,000 samples:

\begin{center}
\begin{tabular}{|c|c|c|c|c|c|}
\hline
& p-value for $Q=8.86$ & $\e[Q]$ & $\e[Q^2]$& $\alpha$&$\beta$\\
\hline
simulation (`true' null) & 0.050 & 3.74 & 20.95 &&\\
chi-square est-df & 0.053 & 3.70 & & &\\
gamma  & 0.037 & 3.70 & 19.37& 2.41 & 1.54\\
chi-square 4 df& 0.065 & 4 & 24&&\\
\hline
\end{tabular}
\end{center}
Notice again that the first and second moments of the `true' null distribution of $Q$ are smaller than the corresponding moments of the chi-square distribution.  Our formulas produce better approximations of these moments, but even with these better approximations the p-value of the approximating gamma distribution is only slightly more accurate than that produced by the chi-square distribution.    The p-value from the chi-square distribution with 3.70 d.f. (0.053) is very close to that of the simulations (0.050). The sample sizes which appear in this meta-analysis (about 10 patients in each of the two arms of the studies) are simply too small for the asymptotics implicit in our formulas for the second moment of $Q$ to be valid. It is somewhat surprising, but gratifying, that the method based on the chi-square distribution with fractional d.f. is so accurate in this example.  For meta-analyses with samples of such small sizes, perhaps the best method of finding a p-value associated with the obtained value of $Q$ is the bootstrap type procedure which we used above: conduct a large simulation with the sample sizes of the actual data and the weighted average of the effects used as a null value.

\subsection{Generalizations from the examples}
There are some features of the examples which are common not only to the two examples but also to all the simulations we have conducted.  We wish to comment on some of these here.  Notice that the mean of the null distribution for $Q$ found via the simulations is somewhat less than the chi-square mean of $I-1$; and the second moment of $Q$ is substantially less than the chi-square second moment of $I^2-1$.  These facts appear to be general.  The formulas of Sections \ref{sec:theory} and \ref{sec:SMD} which we use for estimating the mean and second moment of $Q$ underestimate both the moments but provide estimates which are substantially closer than the chi-square values to the simulated values.   The formula which estimates the mean seems to be very accurate, but the formula for estimating the second moment is not as good.   The over-estimation by the chi-square approximation results, as is well known (see for example \cite{VW-2007a}), in a conservative hypothesis test; that is, the null hypothesis is not rejected often enough.  The underestimation by our formulas results in a slightly liberal hypothesis test when the gamma approximation is used, but in general the p-values are closer to the true values than the chi-square approximation is to the true values. The chi-square with estimated $\e(Q)$ degrees of freedom provides nearly perfect fit.

The fit of the gamma family of distributions to the empirical distribution of $Q$ is remarkably close.  The inaccuracy in the p-values given by our gamma approximation appears to be due to the underestimation of the second moment of $Q$.  In fact, if we were able to accurately estimate the second moment of $Q$, then the estimated p-values would agree with the simulated p-values in our examples to three decimals.  We do not understand the reason why the expansion for $\e[Q^2]$ is not more accurate, or why it always seems to underestimate the second moment.  Resolution of this question is an area of possible future research.

\section{Simulations} \label{sec:SMDsims}

The simulations were performed using the R programming language \citep{rrr}.  The details of the simulations are presented in four tables (Tables~\ref{tab:null2}, \ref{tab:null1}, \ref{tab:null3} and \ref{tab:power}), all of which compare the $Q$ test using the usual chi-square approximation to the $Q$ test using the gamma approximation and the chi-square approximation with fractional degrees of freedom presented in this article. Table~\ref{tab:null2} contains results of the $Q$ test under the null hypothesis in the situation where all studies have the same size, the treatment and control arms are equal, and the combined effect $\delta$ is estimated by $\hat{g}_w$.  Table~\ref{tab:null1} contains results similar to that of Table~\ref{tab:null1}, but here the combined effect is estimated by $\hat{g}_A$.  (See the end of Section~\ref{sec:SMDnotation} for the distinction between $\hat{g}_A$ and $\hat{g}_w$.)  Table~\ref{tab:null3} also contains results of the $Q$ test under the null hypothesis, but in the situation in which the study sizes are not equal.  Finally Table~\ref{tab:power} contains simulation results about the power of the $Q$ test.

\subsection{Simulations under the null hypothesis: equal study sizes}
\label{subsec:simnullequal}

Since $\sqrt{Nq(1-q)}\hat{g}/J\sim t_{N-2}(\sqrt{Nq(1-q)}\delta)$ the values of $\hat{g}$ could be simulated directly from the appropriately scaled non-central $t$-distribution. In this case the quality of simulations would depend on the implementation of the noncentral $t$. Instead we calculated $\hat{g}$ from the first principles, using $\sigma_C=\sigma_T=1$, and simulating sample means $\bar{X}_C\sim N(0,n_C^{-1})$,  $\bar{X}_T\sim N(\delta,n_T^{-1})$ and sample variances $(n_C-1)s^2_C\sim\chi^2_{n_C-1}$ and $(n_T-1)s^2_T\sim\chi^2_{n_T-1}$.\\

The first series of simulations was performed for the situation in which all $I$ of the studies have equal sample sizes.  The data pattern used in the first series of simulations are described in  Tables~\ref{tab:datanull1}. Each data pattern was replicated 100,000 times.  The results of these simulations for the case of equal treatment and control arms ($q=1/2$) appear in Tables~\ref{tab:null1} and ~\ref{tab:null2}.
\begin{table}[ht] \label{tab:datanull1}
  \begin{footnotesize}
   \begin{center}
   \begin{tabular}
   {|l|l|}
     \hline
    $I$ (number of studies)& 5, 10, 20, 50\\
    $N$ (total size of both arms of each study) & 20, 30, 40, 100, 200\\
    $q$ (proportion of each study size in the control arm) & 1/2, 3/4\\
    $\delta$ (null value of the SMD) & 0, 0.2, 0.5, 1, 2\\
\hline
\end{tabular}
\caption{\label{tab:datanull1} \emph{Data pattern of the simulations used in Tables~\ref{tab:null1} and Tables~\ref{tab:null1} \ref{tab:null2} for the Type I error in the $Q$ test }}
\end{center}
\end{footnotesize}
\end{table}

The choice of $\delta$ values was determined by the standard convention \citep{Cohen-1988} that the $\delta$ values of $0.2$ and $0.5$ constitute small and medium effect sizes, respectively. Instead of using the traditional `large' effect size of 0.8, we moved beyond to values of 1 and 2 to explore the possible consequence on the Q test of very large values of $\delta$.   Previous simulations by \cite{VW-2007a} did not uncover any such consequence for $\delta$ values up to $0.8$.\\

 Four p-values were obtained for each value of $Q$ calculated from one of the 100,000 replications: the standard chi-square based p-value; the p-value based on the gamma approximation using the known value of $\delta$ together with the formulas given in Equations~\ref{eq:EQ1} and \ref{eq:EQ2};  the p-value based on the gamma approximation using the estimated null value of $\delta$ together with the formulas given in Equations~\ref{eq:EQ1} and \ref{eq:EQ2}; and the p-value based on the chi-square approximation using the estimated degrees of freedom equal to $\e(Q)$.  These p-values were then compared to the levels $\alpha=0.05$ and $\alpha=0.1$ to obtain the type I errors of each approximation at the  5\% and 10\% nominal levels. In the tables below these values are  denoted by $\chi^2_{\alpha}$,   $\Gamma_{\alpha}^{th}$, $\Gamma_{\alpha}^{s}$, and $\chi^2_{E(Q),\alpha}$ respectively.

In addition to the three p-values ($\chi^2_{\alpha}$, $\Gamma_{\alpha}^{s}$, and $\chi^2_{E(Q),\alpha}$), Table~\ref{tab:null2} contains the first two moments of $Q$ calculated from our formulas with known $\delta$ (denoted $\e_f[Q]$ and $\e_f[Q^2]$ in the table, where the subscript $f$ denotes a result calculated from our approximation formulas) and their sample counterparts $\bar{Q}$ and $\bar{Q^2}$; Table~\ref{tab:null1} additionally provides the fourth p-value $\Gamma_{\alpha}^{th}$, the variance $\var_f[Q]$ and the sample variance $s^2(Q)$.  These data permit us to judge the accuracy of the formulas which give approximations for the moments of $Q$ by comparing the formula values with the simulated distribution of $Q$.\\

\noindent \textbf{\emph{Results of the simulations with equal study sizes }}

The first set of simulations can be used to answer two types of questions:  how accurate are the moments estimated by our formulas---especially compared to the accuracy of the standard chi-square approximation?; and how accurate are the p-values (at the nominal levels 0.05 and 0.10) given by the gamma approximation and the chi-square approximation with fractional degrees of freedom---especially in comparison with the p-values produced by the standard chi-square approximation?  We begin with the moments.\pagebreak

\noindent \emph{Accuracy of the approximating moments}

The simulations provide us with sample estimates of the moments of $Q$ denoted $\bar{Q}$ and $\bar{Q^2}$, which we take to be `true' values.   Thus we can estimate the relative error in the first moment of the two approximations by $(\e_f[Q]/\bar{Q}-1)\times 100\%$ and $((I-1)/\bar{Q}-1)\times 100\%$; and similarly estimate the relative errors in the second moments by $(\e_f[Q^2]/\bar{Q^2}-1)\times 100\%$ and $((I^2-1)/\bar{Q^2}-1)\times 100\%$.

\begin{figure}[h]\vspace{.2cm}\centerline{
\includegraphics [width=5cm,height=6cm]{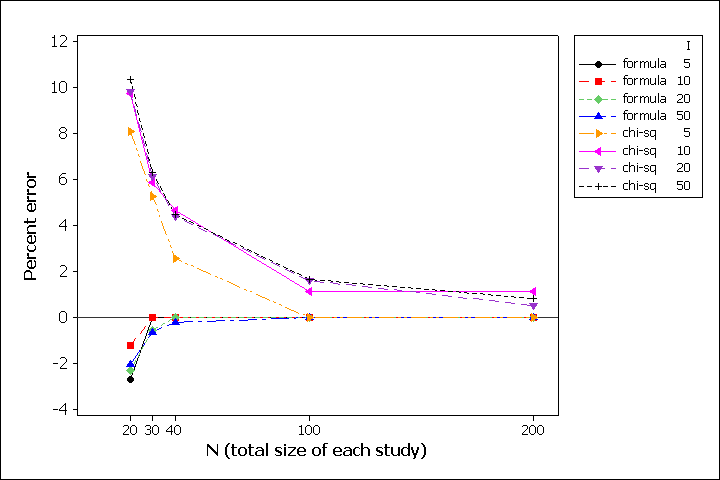}
\includegraphics [width=5cm,height=6cm]{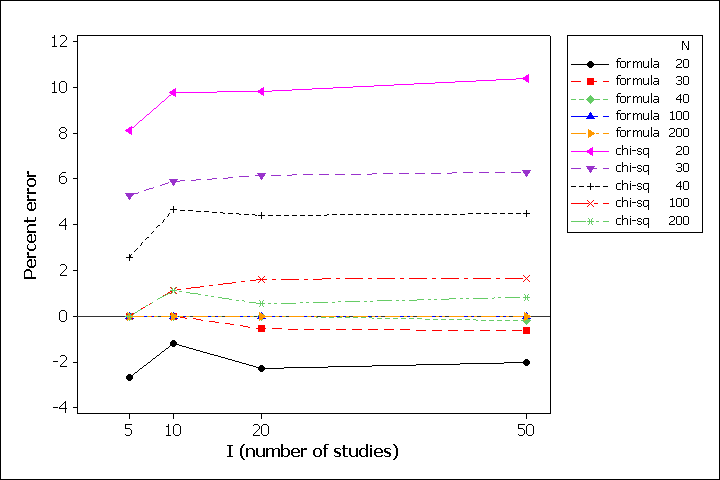}
\includegraphics [width=5cm,height=6cm]{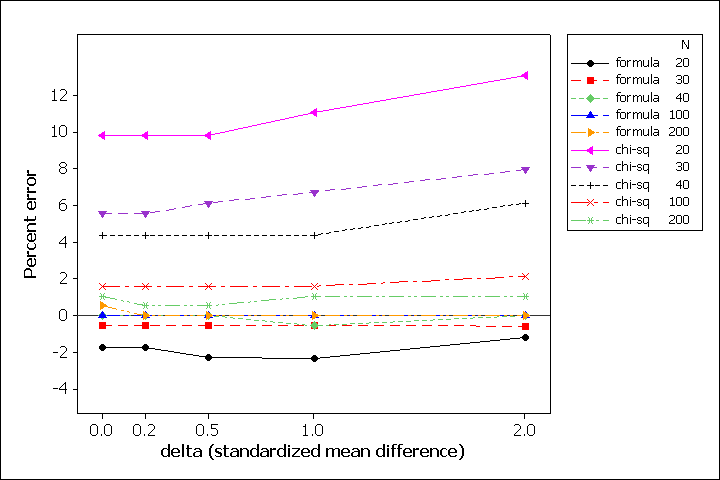}}
\vspace{-.3cm}
\caption{ \label{fig:EQ}\emph{Relative error of two approximations to the mean of $Q$ as a function of the total sample size of each study $N$ (left), of the number of studies $I$  (center), and of the standardized mean difference $\delta$ (right). The lower curves are based on Equation~\ref{eq:EQ1} and the upper curves are from the chi-square first moment.  On the first and the second plots the null value of the SMD $\delta$ is fixed at 0.5. On the rightmost plot,  the number of studies is fixed at $I=20$.}}
\end{figure}

The three graphs of Figure~\ref{fig:EQ} provide a summary of the comparison of the two approximations to the first moment.

We see that $\e_f[Q]$ is generally quite accurate although it slightly underestimates $\bar{Q}$.  In fact the relative error in $E_f[Q]$ is almost always less than 3\%, is less than 1\% for samples of size $N=30$, and is essentially perfect beginning with sample sizes of $N=40$.  In contrast, the chi-square moment is always too large, with relative errors more than  10\% when N=20 and around 5\% when $N=30$ or 40.  Except for the case when the number of studies is small ($I=5$), the relative error of the chi-square first moment remains as high as 1--2\% even when the study sizes are as large as $N=200$.  We also see from the graphs that the relative errors do not seem to depend on the number of studies $I$ or on the standardized mean difference $\delta$, with the exception that for the chi-square approximation the relative error in the first moment increases slightly for the very large (and somewhat unrealistic) values of $\delta=1$ and 2.\\

\begin{figure}[h]\vspace{.2cm}\centerline{
\includegraphics [width=5cm,height=6cm]{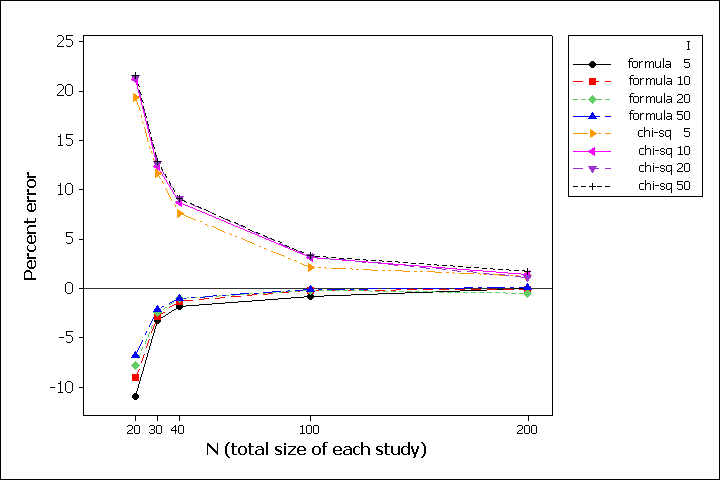}
\includegraphics [width=5cm,height=6cm]{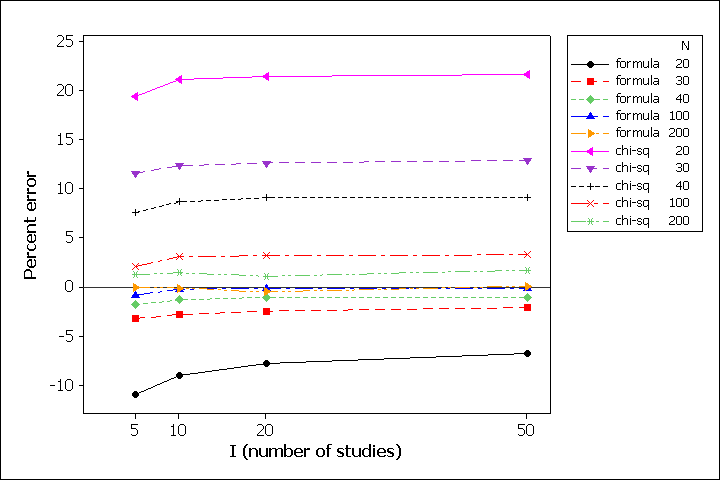}
\includegraphics [width=5cm,height=6cm]{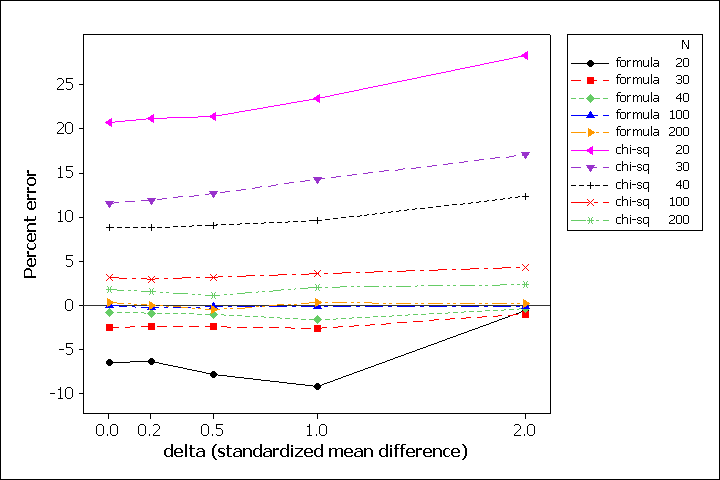}}
\vspace{-.3cm}
\caption{ \label{fig:EQ2}\emph{Relative error of two approximations to the second moment of $Q$ as a function of the total sample size of each study $N$ (left), of the number of studies $I$  (center), and of the standardized mean difference $\delta$ (right). The lower curves are based on Equation~\ref{eq:EQ2} and the upper curves are from the chi-square second moment.  On the first and the second plots the null value of the SMD $\delta$ is fixed at 0.5. On the rightmost plot,  the number of studies is fixed at $I=20$.}}
\end{figure}

The three graphs of Figure~\ref{fig:EQ2} provide a summary of the comparison of the percent error in the approximation of the second moment $\e[Q^2]$ by the two approximating distributions.

We see that the chi-square approximation overestimates the second moment while our formula underestimates the second moment, but by a smaller amount.  The percent error for both approximations decreases as total sample size $N$ increases. The chi-square error starts at about 20\% for $N=20$ and decreases to 9\% for $N=40$ and at $N=100$ the error is still in the 2--3\% range.  In contrast, the error using our formula starts at about 9\% for $N=20$, decreases to less than 2\% for $N=40$ and at $N=100$ the error is less than 1\%.

We see from the graphs that the relative error in the second moment does not appear to have much dependence on the number of studies $I$, except that there is a small difference in error for the very small number of studies $I=5$.
The relative error for the formula values $\e_f[Q^2]$ seems to be independent of $\delta$, but surprisingly there is some increase in the relative error of the chi-square approximation as $\delta$ increases, especially for the very large values of $\delta=1$ and 2.\\


\noindent \emph{Accuracy of significance levels: two-moment gamma vs standard chi-square approximation}

\begin{figure}[h]\vspace{.2cm}\centerline{
\includegraphics [width=7.5cm,height=6cm]{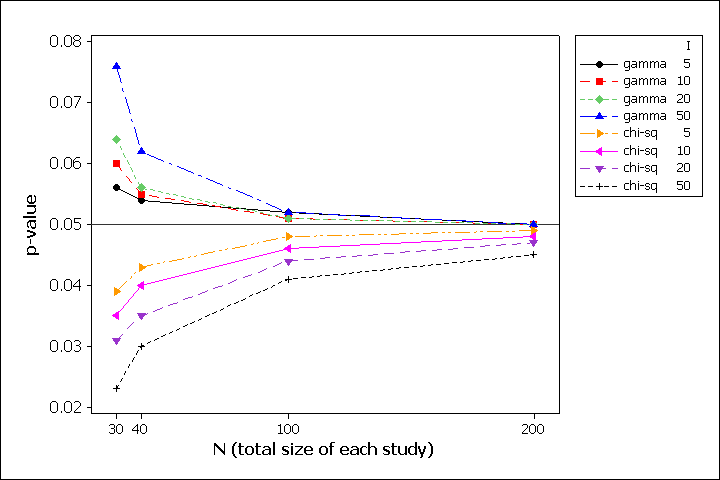}
\includegraphics [width=7.5cm,height=6cm]{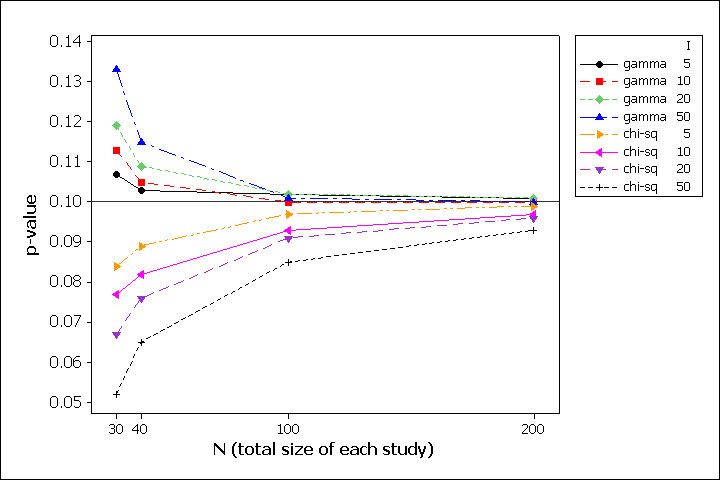}}
\vspace{-.3cm}
\caption{ \label{fig:levelvsN}\emph{Achieved levels of the $Q$ test at the nominal level of 0.05 (left) and 0.1 (right)  using two approximations, as a function of the total sample size of each study $N$. The upper curves are from the gamma approximation and the lower curves are from the chi-square approximation.  The null value of the SMD $\delta$ is fixed at 0.5.  To better show details, the data for $N=20$ have been omitted.}}
\end{figure}

The dependence of the achieved level on the size of the studies $N$ for our gamma and the standard chi-square approximations can be seen graphically in Figure~\ref{fig:levelvsN}.

\begin{figure}[h]\vspace{.2cm}\centerline{
\includegraphics [width=7.5cm,height=6cm]{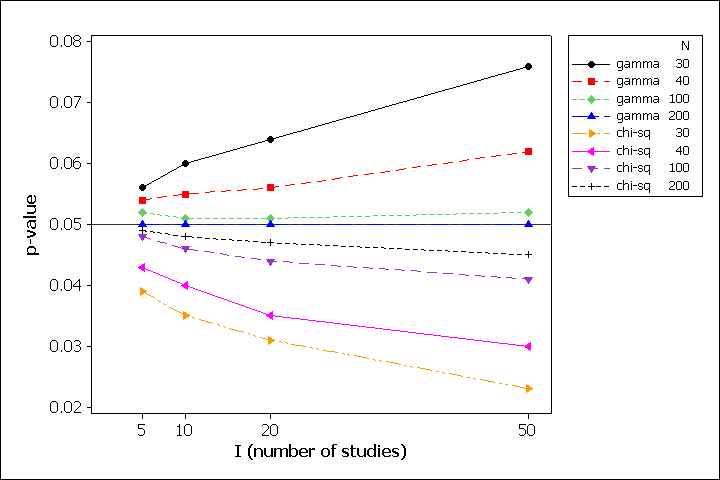}
\includegraphics [width=7.5cm,height=6cm]{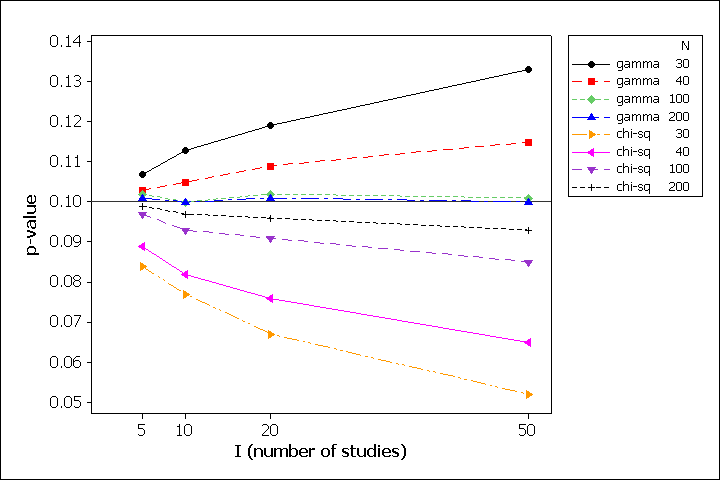}}
\vspace{-.3cm}
\caption{ \label{fig:levelvsI}\emph{Achieved levels of the $Q$ test at the nominal level of 0.05 (left) and 0.1 (right)  using two approximations, as a function of the number of studies $I$. The upper curves are from the gamma approximation and the lower curves are from the chi-square approximation.  The null value of the SMD $\delta$ is fixed at 0.5.  To better show details, the data for $N=20$ have been omitted.}}
\end{figure}

\begin{figure}[h]\vspace{.2cm}\centerline{
\includegraphics [width=7.5cm,height=6cm]{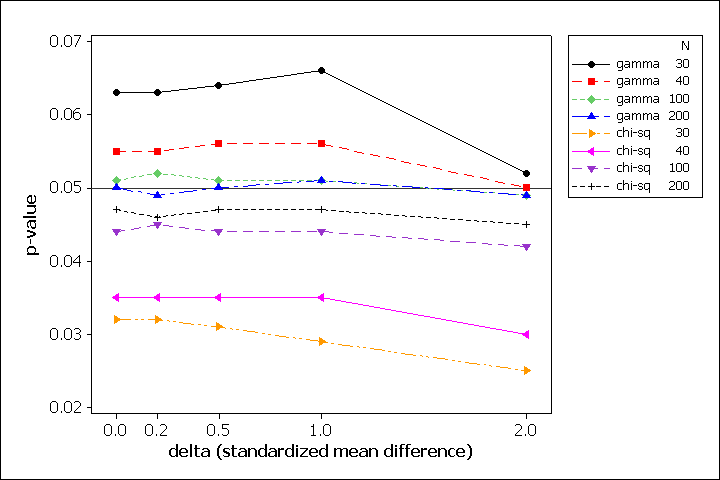}
\includegraphics [width=7.5cm,height=6cm]{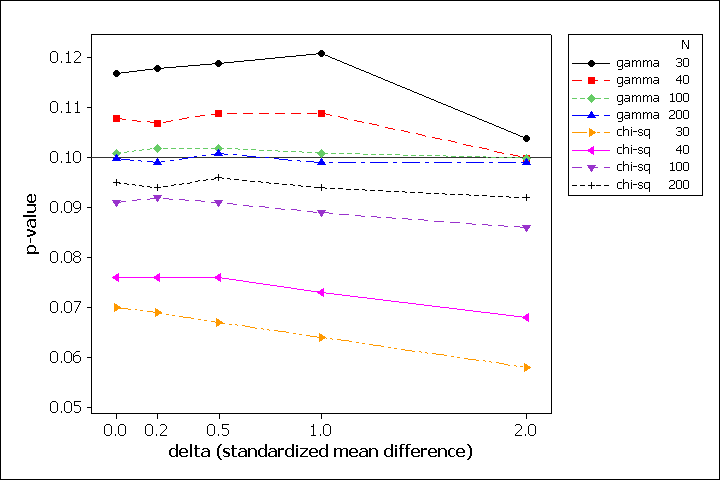}}
\vspace{-.3cm}
\caption{ \label{fig:levelvsdelta}\emph{Achieved levels of the $Q$ test at the nominal level of 0.05 (left) and 0.1 (right)  using two approximations, as a function of the standardized mean difference $\delta$. The upper curves are from the gamma approximation and the lower curves are from the chi-square approximation.  The number of studies is fixed at $I=20$.  To better show details, the data for $N=20$ have been omitted.}}
\end{figure}

The type I error of the $Q$ test of homogeneity using the standard chi-square approximation is considerably lower than the nominal level, and hence the standard test is very conservative.  This conservativeness is a well known fact; our simulations agree with the simulations of \cite{sanches-1997}, \cite{VW-2007a}, and others.  Because the standard test is so conservative, there is a well known recommendation to use the 10\% significance level for the $Q$ test (see \cite{pet-2001}, among others).  Our simulations confirm that this recommendation is certainly justified; for 10 or more small studies ($N=20$), the type I error at the 10\% significance level is closer to 5\% than to 10\%.

In contrast, our gamma approximation is somewhat liberal for small values of $N$. In fact, for total study sizes as small as $N=20$ the gamma approximation is sufficiently poor that we do not recommend it.  For $N=30$ the true level seems to be in between the  two approximations. Starting from $N=40$ the gamma approximation works better than the standard chi-square approximation. For a fixed value of $I$, the performance of both approximations improves with the study size, but the improvement is considerably faster for the gamma approximation.  For $N=100$ the gamma approximation delivers perfect results, whereas the chi-square approximation is still too conservative.

For fixed study size $N$, the accuracy of the achieved levels decays as the number of studies $I$ increases.  For example, for the gamma approximation, studies of size 40 (and even size 30) provide reasonably accurate levels when there are only $I=5$ studies.  However when the number of studies increases to $I=50$, then larger study sizes are necessary to achieve accurate levels.  For $I=50$, studies of size 40 are not large enough, but studies of size 100 give excellent results.  For an intermediate number of $I=20$ studies, the study size of $N=40$ gives reasonably accurate levels producing levels of about 0.055 and 0.108 for nominal levels of 5\% and 10\% respectively.  The pattern is similar for the chi-square approximation: meta-analyses with many studies require large sample sizes for accuracy.  But in all cases, the chi-square performs less well than the gamma approximation.  The dependence of the behavior of the achieved levels on $I$ can be seen in Figure~\ref{fig:levelvsI}.

The simulations show that the type I error of the standard chi-square test decreases as the effect size $\delta$ increases. Thus the test is even more conservative for larger effect sizes.  However, the gamma approximation improves as the effect size $\delta$ increases, contrasting with the worsening of the chi-square approximation.  The dependence of the behavior of the achieved levels on $\delta$ can be seen in Figure~\ref{fig:levelvsdelta}.\pagebreak

\noindent \emph{Accuracy of significance levels for the chi-square approximation with fractional degrees of freedom}

The results of simulations to do with the fractional chi-square test are not included in the figures. As can be seen from Table~\ref{tab:null2}, in every instance, the fractional chi-square test is superior to the usual chi-square test.  Most importantly for applications is the fact that the improvement given by the fractional chi-square is substantial for small to moderate sample sizes, from $N=20$.  As examples of this improvement, consider the case of $I=20$ studies and $\delta=0.5$.  The simulations indicate the following improvements in the achieved level at the two nominal levels of 0.05 and 0.10: for $N=20$ the achieved levels improve from 0.021 to 0.046 and from 0.050 to 0.098, respectively; for $N=40$ from 0.035 to 0.047 and from 0.076 to 0.096, respectively; and even for study size as large as $N=100$, the achieved levels improve from 0.044 to 0.048 and from 0.090 to 0.099, respectively.\\

\noindent \emph{Other results of the equal study size simulations}

First, the simulations of Table~\ref{tab:null2} were repeated with equal total study sizes as before, but with each study having an unbalanced design with three-quarters of the study size present in the control arm ($q=3/4$).  The results were so similar to that of the balanced studies that we have not included either a table of the results analogous to Table~\ref{tab:null2} nor graphical displays of the data.

Second, there is not much difference between the type I error with a known value of $\delta$ (denoted by $\Gamma_{\alpha}^{th}$ in Table~\ref{tab:null1}) and the type I error with an estimated null value of $\delta$ (denoted by $\Gamma_{\alpha}^{s}$ in Tables~\ref{tab:null2} and \ref{tab:null1}). Of course, only the latter test can be used in practice.

Finally, the results in Table~\ref{tab:null2} used the estimated null value of  $\hat{\delta}=\sum w_i\delta_i/\sum w_i$. In Table~\ref{tab:null1} the simulations were repeated for using  $\hat{\delta}=\sum A_i^{-1}\delta_i/\sum A_i^{-1}$ instead. It is known \citep{Yuan} that the former, more natural, estimator is a biased estimator of the combined null value of $\delta$. Does the choice of estimator of $\delta$ affect the results? It can be seen that the only noticeable differences are for $N=20$ and $\delta=2$. Then the constant weights $A_i^{-1} $ provide p-values closer to those obtained using the known value of $\delta$ when using gamma approximation. Interestingly, the inverse variance weights provide p-values closer to nominal for $K=10$ and $K=20$, but not for $K=5$ or $K=50$. These differences are only academic though, we do not recommend our gamma approximation for $N=20$ in any case, and $\delta=2$ is much too large. Thus, there is no practical  difference between the two choices, take your pick.

\subsection{Simulations under the null hypothesis: unequal study sizes}

The second series of simulations used unequal study sizes.  We have followed a suggestion of  \cite{sanches-2000}, who selected the following study sizes with  the skewness of 1.464 which they consider typical for meta-analyses in the field of behavioral and health sciences:
the set $N_1$ with average study of sixty, consisting of individual sizes $\{24,32,36,40,168\}$;
the set $N_2$ with average study size of 100, consisting of individual sizes $\{64,72,76,80,208\}$ and the set  $N_3$ with average study size of 160, consisting of individual study sizes $\{124,132,136,140,268\}$. We have taken the studies to be balanced, thus dividing each study size equally between the two study arms.
The simulations were run for $I=5,\;10$ and $20$. For  meta-analyses with $I=10$ and $I=20$,  the same set of sample sizes was repeated twice or four times, respectively.  The data patterns of the simulations are summarized in Table~\ref{tab:datanull2}.

\begin{table}[ht]
  \begin{footnotesize}
   \begin{center}
   \begin{tabular}
   {|l|l|}
     \hline
    $I$ (number of studies)& 5, 10, 20\\
    $\bar{N}$ (average and (individual) study sizes) & 60 (24, 32, 36, 40, 168)\\ & 100 (64, 72, 76, 80, 208)\\
    & 160 (124, 132, 136, 140, 268) \\
    $q$ (proportion of each study in the control arm) & 1/2\\
    $\delta$ (null value of the SMD) & 0.5\\
\hline
\end{tabular}
\caption{\label{tab:datanull2} \emph{Data pattern of the simulations used in Table~\ref{tab:null3} for the Type I error in the $Q$ test for unequal study sizes}}
\end{center}
\end{footnotesize}
\end{table}

\bigskip
\noindent \textbf{\emph{Results of the simulations with unequal study sizes }}

The results of the simulations with unequal study sizes are given in Table~\ref{tab:null3}.  The approximation of the moments is excellent.  The first moments given by the formulas are nearly exact (relative error less than 2\%) and the second moments have relative error less than 3\%, compared with relative errors of the chi-square first and second moments of more than 5\% and 10\% respectively.

\begin{figure}[h]\vspace{.2cm}\centerline{
\includegraphics [width=7.5cm,height=6cm]{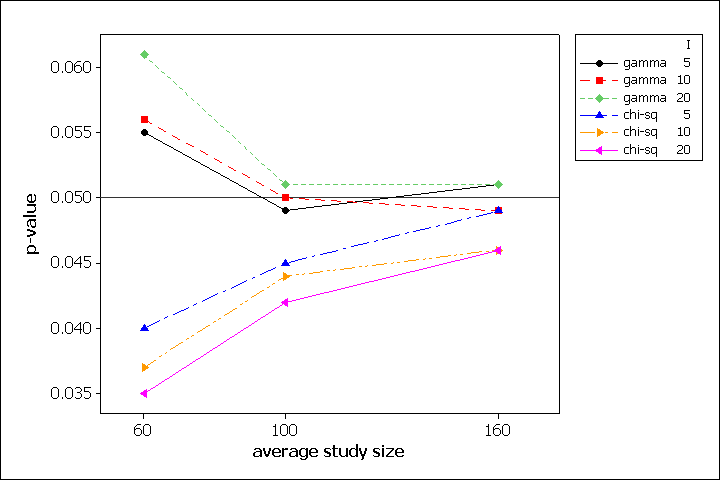}
\includegraphics [width=7.5cm,height=6cm]{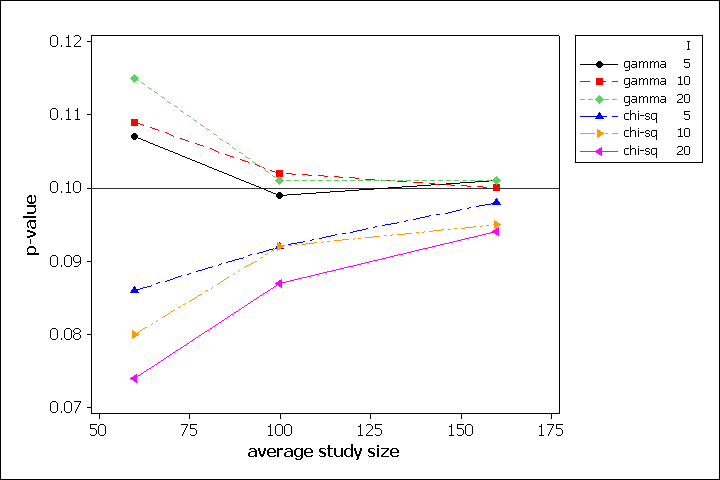}}
\vspace{-.3cm}
\caption{ \label{fig:level.uneq}\emph{Achieved levels of the $Q$ test at the nominal levels of 0.05 (left) and 0.1 (right)  using two approximations, as a function of the average sample size of each study $\bar{N}$. The sample sizes are unequal in this figure.  The upper curves are from the gamma approximation and the lower curves are from the chi-square approximation.  $I$ is the number of studies.  The standardized mean difference has been fixed at $\delta=0.5$.}}
\end{figure}

The significance levels are similar to those obtained from the simulations for equal study sizes.  The chi-square approximation yields a conservative test, while the gamma approximation yields a liberal test which is closer to the nominal levels.   At the significance levels of 0.05 and 0.10, the gamma approximation is nearly perfect for the larger two sizes $\bar{N}=100$ and 160 while the error in the level of the chi-square approximation is substantial even for the largest size of $\bar{N}=160$.  For the smaller size of $\bar{N}=60$, the gamma approximation has an error of roughly half that of the chi-square approximation.  Graphical displays of the levels are shown in Figure~\ref{fig:level.uneq}. Once more, the results from the fractional chi-squre approximation are nearly perfect even for the smallest sample sizes.

\subsection{Comparison of the power of the Q tests}

The standard $Q$ test using the chi-square approximation is well known to have low power (see for example \cite{VW-2007a}).  In this section we report on simulations to see how the power of the $Q$ test is improved by the use of  our moment approximations.  To this end, we adopt the random effects model that the heterogeneity in effects among the several studies is modeled by the assumption that the effect $\delta_i$ of the $i$th study is normally distributed about a fixed mean $\delta$ and with variance $\tau^2$.  Then the null (homogeneity) hypothesis becomes $\tau^2=0$ and alternatives are measured by the magnitude of $\tau^2$.  In the simulations, we have taken $\delta=0.5$, a `medium' effect size, and have varied $\tau^2$ from 0.025 to 0.25.  We compared the power of the standard and the improved tests in the range from $N=20$ to $N=80$ where we expected noticeable differences: in general, the power of the standard $Q$ test is considered not to be sufficient for $N\leq 80$ (\cite{VW-2007a}).  The data patterns for the power simulations are specified in Table~\ref{tab:altdata}.
\begin{table}[ht]
  \begin{footnotesize}
   \begin{center}
   \begin{tabular}
   {|l|l|}
     \hline
    $I$ (number of studies)& 5, 10, 20, 50\\
    $N$  (equal study sizes) & 20, 30, 40, 50, 60, 80 \\
    $q$ (proportion of each study in the control arm) & 1/2\\
    $\delta$ (null value of the SMD) & 0.5\\
    $\tau^2$ (variance of random effect)&0.025, 0.05, 0.1, 0.15, 0.20, 0.25\\
    $\alpha$ (nominal significance level of the test) & 0.05, 0.10\\
\hline
\end{tabular}
\caption{\label{tab:altdata} \emph{Data pattern of the simulations used in Table~\ref{tab:power} for the power of the $Q$ test.}}
\end{center}
\end{footnotesize}
\end{table}

We conducted 10,000 repetitions for each configuration. We simulated within-study parameters $\delta_i\sim \textbf{N}(\delta,\tau^2)$, $i=1,\cdots, I$ and then simulated the values of $\hat{g} _i$ directly from the appropriately scaled non-central $t$-distribution with non-centrality parameter $\sqrt{Nq(1-q)}\delta_i$.  The results of the simulations appear in Table~ \ref{tab:power}.\\

\noindent \textbf{\emph{Results of the power simulations}}

Since the test based on the gamma approximation is liberal, its power is higher than the power of the conservative standard test.   We note that the power of the test using the fractional chi-square distribution is also always higher than the test using the standard chi-square approximation.  In this discussion, we focus on the magnitude of the improvement in power rather than on the power for the  tests separately. The most striking result of the simulations is that the power \emph{improvement} increases as the number of studies increases and as the sizes of the studies decrease.  The greatest improvement in power for the fractional chi-square test in comparison to the standard test (based on the range of our simulations) is 21  percentage points which occurred for $I=50$ studies, study sizes $N=20$, and for $\tau^2=0.1$.  Maximum improvement for the other values of $I$ were 12 percentage points for $I=20$, 7 percentage points for $I=10$ and 4 percentage points for $I=5$, all occurring at the smallest study size of $N=20$.  As the study sizes $N$ increase from $N=20$ to $N=40$, the improvement in power for the fractional chi-square test decreases by roughly two-thirds.  Finally we note that the increase in power at the two different levels of 0.05 and 0.10 were quite similar to each other.

 Since the gamma approximation is recommended only for $N\geq 40$,  we consider this range when comparing the power of the test based on gamma approximation to the standard chi-square test. The greatest improvement in power is 11 to 12 percentage points which occurred for the largest number of studies $I=50$ and the smallest study sizes $N=40$.  Maximum improvement for the other values of $I$ were 7 percentage points for $I=20$, 5 percentage points for $I=10$ and 3 percentage points for $I=5$.  As the study sizes $N$ increased from $N=40$ to $N=80$, the improvement decreased by roughly half.  Once more,  the increase in power at the two different levels of 0.05 and 0.10 were quite similar to each other.

\section{Summary and concluding discussion}\label{sec:conclude}

The main focus of this paper is the improvement of the test for homogeneity commonly used in meta-analysis by referring Cochran's $Q$ statistic to a more accurate distribution.  In this paper, we have considered the situation in which the $Q$ statistic is a function of only one parameter and have applied the results to the case in which the effect of interest is measured by the standardized mean difference (SMD or Cohen's $d$ statistic), a measure which is frequently used in meta-analytic applications.   We have presented expansions for the first two moments of $Q$ which are accurate to order $O(1/n)$.  These expansions thus offer corrections of order $O(1/n)$ to the corresponding moments of the chi-square approximation to the distribution of $Q$.  These expansions are the first that we are aware of to include the situation in which the weights in the $Q$ statistic are not independent of the effects (as is the case with the SMD).

We considered two options to approximate the distribution of $Q$ for the SMD: the use of a gamma distribution with moments matching those of the expansions or by the chi-square distribution with fractional degrees of freedom matching the first moment.  Both approximations result in  improved $Q$ tests for homogeneity when the effects are measured by the SMD.  To facilitate the substantial computations necessary for these improved tests, a computer program in the \emph{R}-language can be downloaded from  \textbf{http://www.imperial.ac.uk/stathelp/researchprojects/metaanalysis}.

Our simulations show that the improved test for the SMD using the gamma distribution is somewhat liberal (rejecting the null hypothesis more often than appropriate); in contrast, the currently used test which uses the chi-square distribution is well known to be conservative.  But the improved test based on gamma approximation is quite accurate for study sizes of 40 or more (for example, 20 subjects in each arm of a randomized clinical trial).
 
 However our recommended test is based not on the gamma approximation but on the use of the fractional chi-square  distribution whose first moment matches that of the expansion.    In applications, the parameter in the expansion will need to be estimated from the data.  Thus our recommended approximating distribution of $Q$ (namely $\chi^2_{E[Q]}$) is data dependent as opposed to the now standard approximating distribution of $Q$ (namely $\chi^2_{I-1}$) which is data independent.  The result is an improved $Q$ test for homogeneity when the effects are measured by the SMD.  

Simulations show that our recommended improved $Q$ test for homogeneity  yields a substantial improvement over the standard test in accuracy of achieved significance levels, especially for small to moderate study sizes.   In addition the improved test provides an increase in power. The simulations show that the improved test works quite well in a variety of circumstances, such as when the individual studies have unbalanced sizes between the two arms or when the studies have substantially different total sizes from each other.

An important limitation of this paper, which is intended to be the first in a series, is the restriction to the one parameter case.  In future work, we plan to extend our expansions to the two parameter case and to provide applications to important meta-analytic measures such as the risk difference and the odds ratio.\\

\noindent{\bf Acknowledgements}\\
The authors wish to thank Joanne McKenzie for providing data from The Cochrane Database of Systematic Reviews (2004) in an electronic format.

\pagebreak
\appendix
\section{Appendix} \label{sec:Appendix}
Equation~\ref{eq:EQ2} which approximates the second moment of $Q$ needs expressions for various derivatives of $Q^2$ with respect to $\theta$.  These derivatives are provided below. But for ease of reference, we first reproduce Equation~\ref{eq:EQ2}.

\begin{eqnarray*}
\e[Q^2] &\approx& \frac{1}{24}\sum_i \frac{\d^4[Q^2]}{\d\theta_i^4}\; \e[\Theta_i^4]
+ \frac{1}{8}\sum_{i\neq j}\sum\frac{\d^4[Q^2]}{\d\theta_i^2\d\theta_j^2}\;
\e[\Theta_i^2]\e[\Theta_j^2]
+ \frac{1}{120}\sum_i \frac{\d^5[Q^2]}{\d\theta_i^5}\;\e[\Theta_i^5] \\ &+&
\frac{1}{12}\sum_{i \neq j}\sum \frac{\d^5[Q^2]}{\d\theta_i^3\d\theta_j^2} \;
\e[\Theta_i^3]\e[\Theta_j^2]
+\frac{1}{720}\sum_i \frac{\d^6[Q^2]}{\d\theta_i^6}\;\e[\Theta_i^6] \\
&+&\frac{1}{48} \sum_{i \neq j}\sum  \frac{\d^6[Q^2]}{\d\theta_i^4\d\theta_j^2}  \;
\e[\Theta_i^4]\e[\Theta_j^2]
 +\frac{1}{48}\sum_{i \neq j}\sum_{\neq
k}\sum\frac{\d^6[Q^2]}{\d\theta_i^2\d\theta_j^2\d\theta_k^2} \;
\e[\Theta_i^2]\e[\Theta_j^2]\e[\Theta_k^2]
\end{eqnarray*}

Here are the derivatives of $Q^2$ needed for the above formula.

\begin{eqnarray}
\frac{\d^4[Q^2]}{\d\theta_i^4} &=& 24 w_i^2U_i^2 \label{eq:d^4(Q^2)}\\
\frac{\d^4[Q^2]}{\d\theta_i^2\d\theta_j^2} &=& 8w_iw_j\left(U_iU_j+\frac{2w_iw_j}{W^2}\right) \label{eq:d^4(Q^2ij)}\\
\frac{\d^5[Q^2]}{\d\theta_i^5} &=& 240 w_iU_i^3 \frac{d f_i}{d\theta_i}\\
\frac{\d^5[Q^2]}{\d\theta_i^3\d\theta_j^2} &=& 24 U_iw_j\left[U_iU_j +\frac{ 5w_iw_j}{W^2}\right] \frac{d f_i}{d\theta_i}
 -48 \frac{w_i^2}{W}\left[ U_iU_j+\frac{w_iw_j}{W^2}\right] \frac{d f_j}{d\theta_j} \\
 \frac{\d^6[Q^2]}{\d\theta_i^6} &=& 720U_i^3\left[\left(U_i-\frac{2w_i}{W}\right)\left(\frac{d f_i}{d\theta_i}\right)^2 + w_i \frac{d^2 f_i}{d\theta_i^2}\right]\\
 \frac{\d^6[Q^2]}{\d\theta_i^4\d\theta_j^2} &=& \frac{48}{W^4}\left\{ -2WU_iw_j(W^2U_i-4Ww_j+9w_iw_j)\left(\frac{d f_i}{d\theta_i}\right)^2 \right. \\&&+ W^2U_iw_j(W^2-Ww_j-Ww_i+6w_iw_j)\frac{d^2 f_i}{d\theta_i^2} \nonumber\\
&&- 8WU_iw_i(W^2U_i-Ww_j+3w_iw_j)\left(\frac{d f_i}{d\theta_i}\right)\left(\frac{d f_j}{d\theta_j}\right) \nonumber\\
&&+  \left. w_i^3(-2W+3w_i)\left(\frac{d f_j}{d\theta_j}\right)^2 +W^2U_iw_i^3 \frac{d^2 f_j}{d\theta_j^2}\; \right\} \nonumber\\
\frac{\d^6[Q^2]}{\d\theta_i^2\d\theta_j^2\d\theta_k^2} &=& -\frac{16w_jw_k}{W^4} (Ww_j+Ww_k-9w_jw_k) \left(\frac{d f_i}{d\theta_i}\right)^2\\
&&- \frac{16w_iw_k}{W^4} (Ww_i+Ww_k-9w_iw_k) \left(\frac{d f_j}{d\theta_j}\right)^2 \nonumber \\
&&- \frac{16w_iw_j}{W^4} (Ww_i+Ww_j-9w_iw_j) \left(\frac{d f_k}{d\theta_k}\right)^2  \nonumber \\
&& + \frac{8w_jw_k}{W^3} (Ww_j+Ww_k-6w_jw_k) \frac{d^2 f_i}{d\theta_i^2}
\nonumber \\
&& + \frac{8w_iw_k}{W^3} (Ww_i+Ww_k-6w_iw_k) \frac{d^2 f_j}{d\theta_j^2}
\nonumber \\
&& + \frac{8w_iw_j}{W^3} (Ww_i+Ww_j-6w_iw_j) \frac{d^2 f_k}{d\theta_k^2}
\nonumber \\
&&-\frac{32w_k}{W^2}\left[U_iU_j(W-6w_k)+\frac{w_iw_j}{W^2}(W-12w_k) +3w_k\right] \left(\frac{d f_i}{d\theta_i}\right)\left(\frac{d f_j}{d\theta_j}\right) \nonumber\\
&&-\frac{32w_j}{W^2}\left[U_iU_k(W-6w_j)+\frac{w_iw_k}{W^2}(W-12w_j) +3w_j\right] \left(\frac{d f_i}{d\theta_i}\right)\left(\frac{d f_k}{d\theta_k}\right) \nonumber\\
&&-\frac{32w_i}{W^2}\left[U_jU_k(W-6w_i)+\frac{w_jw_k}{W^2}(W-12w_i) +3w_i\right] \left(\frac{d f_j}{d\theta_j}\right)\left(\frac{d f_k}{d\theta_k}\right) \nonumber
\end{eqnarray}

\pagebreak
\bibliography{KMS1}

\newpage
      \begin{footnotesize}
   \begin{longtable} {|c|c|c||r|r|r||r|r|r||r|r|r|r|}
     \hline
$I$ &   $N$ &   $\delta$    &   $\chi^2_{.05}$  &   $\Gamma_{.05}^{s}$ &   $\chi^2_{E(Q),.05}$  &    $\chi^2_{.1}$  &    $\Gamma_{.1}^{s}$ &   $\chi^2_{E(Q),.1}$   &   $E_f(Q)$  &   $\bar{Q}$   &   $E_f(Q^2)$     &  $\bar{Q^2}$  \\
\hline
5	&	10	&	0	&	0.014	&	NA	&	0.052	&	0.041	&	NA	&	0.119	&	2.8	&	3.2	&	-12.2	&	14.9	\\
5	&	10	&	0.2	&	0.014	&	NA	&	0.051	&	0.039	&	NA	&	0.119	&	2.7	&	3.2	&	-14.3	&	14.8	\\
5	&	10	&	0.5	&	0.013	&	NA	&	0.052	&	0.037	&	NA	&	0.120	&	2.6	&	3.1	&	-22.7	&	14.5	\\
5	&	10	&	1	&	0.012	&	NA	&	0.055	&	0.035	&	NA	&	0.127	&	2.5	&	3.1	&	-24.4	&	14.1	\\
5	&	10	&	2	&	0.008	&	NA	&	0.039	&	0.027	&	NA	&	0.097	&	2.8	&	3.0	&	134.3	&	12.8	\\
\hline																									
5	&	14	&	0	&	0.026	&	NA	&	0.044	&	0.061	&	NA	&	0.097	&	3.4	&	3.5	&	12.7	&	18.1	\\
5	&	14	&	0.2	&	0.026	&	NA	&	0.044	&	0.062	&	NA	&	0.097	&	3.4	&	3.5	&	12.4	&	18.0	\\
5	&	14	&	0.5	&	0.025	&	NA	&	0.045	&	0.061	&	NA	&	0.098	&	3.3	&	3.5	&	10.9	&	17.9	\\
5	&	14	&	1	&	0.023	&	NA	&	0.043	&	0.056	&	NA	&	0.097	&	3.2	&	3.4	&	8.9	&	17.3	\\
5	&	14	&	2	&	0.019	&	NA	&	0.040	&	0.049	&	NA	&	0.093	&	3.2	&	3.3	&	20.0	&	16.3	\\
\hline																									
5	&	16	&	0	&	0.030	&	0.105	&	0.045	&	0.068	&	0.161	&	0.096	&	3.5	&	3.6	&	15.7	&	19.1	 \\
5	&	16	&	0.2	&	0.030	&	0.110	&	0.044	&	0.068	&	0.164	&	0.097	&	3.5	&	3.6	&	15.4	&	19.0	 \\
5	&	16	&	0.5	&	0.029	&	0.124	&	0.044	&	0.066	&	0.178	&	0.095	&	3.5	&	3.6	&	14.5	&	18.8	 \\
5	&	16	&	1	&	0.028	&	0.160	&	0.044	&	0.063	&	0.212	&	0.097	&	3.4	&	3.5	&	13.0	&	18.3	 \\
5	&	16	&	2	&	0.022	&	0.061	&	0.040	&	0.055	&	0.112	&	0.091	&	3.3	&	3.4	&	17.8	&	 17.2\\
\hline																									
5	&	20	&	0	&	0.034	&	0.069	&	0.045	&	0.075	&	0.123	&	0.096	&	3.7	&	3.7	&	18.5	&	 20.3\\
5	&	20	&	0.2	&	0.034	&	0.070	&	0.045	&	0.075	&	0.123	&	0.096	&	3.6	&	3.7	&	18.4	&	20.2	 \\
5	&	20	&	0.5	&	0.034	&	0.073	&	0.046	&	0.074	&	0.127	&	0.095	&	3.6	&	3.7	&	17.9	&	20.1	 \\
5	&	20	&	1	&	0.032	&	0.081	&	0.045	&	0.072	&	0.134	&	0.095	&	3.6	&	3.6	&	16.9	&	 19.7\\
5	&	20	&	2	&	0.028	&	0.058	&	0.042	&	0.064	&	0.110	&	0.092	&	3.5	&	3.6	&	18.1	&	18.7	 \\
\hline																									
5	&	30	&	0	&	0.040	&	0.055	&	0.047	&	0.084	&	0.106	&	0.096	&	3.8	&	3.8	&	21.1	&	21.6	 \\
5	&	30	&	0.2	&	0.039	&	0.054	&	0.046	&	0.084	&	0.106	&	0.096	&	3.8	&	3.8	&	21.0	&	21.5	 \\
5	&	30	&	0.5	&	0.040	&	0.056	&	0.047	&	0.084	&	0.108	&	0.096	&	3.8	&	3.8	&	20.8	&	21.5	 \\
5	&	30	&	1	&	0.038	&	0.058	&	0.046	&	0.082	&	0.109	&	0.096	&	3.8	&	3.8	&	20.3	&	21.2	 \\
5	&	30	&	2	&	0.036	&	0.054	&	0.045	&	0.078	&	0.105	&	0.095	&	3.7	&	3.7	&	20.1	&	20.7	 \\
\hline																									
5	&	40	&	0	&	0.042	&	0.051	&	0.046	&	0.088	&	0.102	&	0.096	&	3.9	&	3.8	&	22.0	&	22.0	 \\
5	&	40	&	0.2	&	0.044	&	0.053	&	0.048	&	0.090	&	0.104	&	0.098	&	3.9	&	3.9	&	22.0	&	22.4	 \\
5	&	40	&	0.5	&	0.043	&	0.054	&	0.048	&	0.088	&	0.103	&	0.097	&	3.8	&	3.9	&	21.9	&	22.3	 \\
5	&	40	&	1	&	0.041	&	0.054	&	0.047	&	0.087	&	0.104	&	0.096	&	3.8	&	3.8	&	21.5	&	21.9	 \\
5	&	40	&	2	&	0.039	&	0.052	&	0.046	&	0.082	&	0.102	&	0.094	&	3.8	&	3.8	&	21.2	&	21.4	 \\
\hline																									
5	&	100	&	0	&	0.048	&	0.051	&	0.050	&	0.097	&	0.101	&	0.100	&	3.9	&	4.0	&	23.3	&	23.4	 \\
5	&	100	&	0.2	&	0.048	&	0.051	&	0.049	&	0.096	&	0.100	&	0.099	&	3.9	&	3.9	&	23.3	&	23.4	 \\
5	&	100	&	0.5	&	0.046	&	0.050	&	0.048	&	0.095	&	0.100	&	0.098	&	3.9	&	3.9	&	23.3	&	23.3	 \\
5	&	100	&	1	&	0.046	&	0.050	&	0.048	&	0.095	&	0.100	&	0.098	&	3.9	&	3.9	&	23.2	&	23.1	 \\
5	&	100	&	2	&	0.045	&	0.050	&	0.048	&	0.093	&	0.100	&	0.098	&	3.9	&	3.9	&	23.0	&	23.0	 \\
\hline																									
5	&	200	&	0	&	0.049	&	0.051	&	0.050	&	0.098	&	0.101	&	0.100	&	4.0	&	4.0	&	23.7	&	23.8	 \\
5	&	200	&	0.2	&	0.049	&	0.050	&	0.050	&	0.097	&	0.100	&	0.099	&	4.0	&	4.0	&	23.7	&	23.6	 \\
5	&	200	&	0.5	&	0.048	&	0.049	&	0.049	&	0.097	&	0.099	&	0.099	&	4.0	&	4.0	&	23.7	&	23.6	 \\
5	&	200	&	1	&	0.049	&	0.050	&	0.050	&	0.097	&	0.100	&	0.099	&	4.0	&	4.0	&	23.6	&	23.7	 \\
5	&	200	&	2	&	0.048	&	0.050	&	0.049	&	0.098	&	0.102	&	0.101	&	4.0	&	4.0	&	23.5	&	23.7	 \\
\hline																									
10	&	10	&	0	&	0.007	&	NA	&	0.068	&	0.022	&	NA	&	0.149	&	5.8	&	7.1	&	-29.7	&	60.2	\\
10	&	10	&	0.2	&	0.007	&	NA	&	0.070	&	0.022	&	NA	&	0.152	&	5.8	&	7.1	&	-35.7	&	60.4	\\
10	&	10	&	0.5	&	0.007	&	NA	&	0.072	&	0.022	&	NA	&	0.157	&	5.6	&	7.0	&	-58.3	&	59.4	\\
10	&	10	&	1	&	0.006	&	NA	&	0.073	&	0.019	&	NA	&	0.162	&	5.4	&	6.9	&	-46.2	&	57.2	\\
10	&	10	&	2	&	0.004	&	NA	&	0.034	&	0.013	&	NA	&	0.085	&	6.9	&	6.6	&	523.3	&	52.9	\\
\hline																									
10	&	14	&	0	&	0.019	&	NA	&	0.047	&	0.047	&	NA	&	0.101	&	7.4	&	7.8	&	54.2	&	73.6	\\
10	&	14	&	0.2	&	0.019	&	NA	&	0.046	&	0.046	&	NA	&	0.102	&	7.4	&	7.8	&	53.2	&	73.7	\\
10	&	14	&	0.5	&	0.018	&	NA	&	0.047	&	0.045	&	NA	&	0.103	&	7.3	&	7.8	&	48.9	&	72.9	\\
10	&	14	&	1	&	0.016	&	NA	&	0.047	&	0.041	&	NA	&	0.104	&	7.2	&	7.7	&	44.1	&	71.2	\\
10	&	14	&	2	&	0.011	&	NA	&	0.037	&	0.032	&	NA	&	0.085	&	7.3	&	7.4	&	88.7	&	66.5	\\
\hline																									
10	&	16	&	0	&	0.022	&	0.156	&	0.044	&	0.053	&	0.214	&	0.096	&	7.8	&	8.0	&	65.1	&	77.3	 \\
10	&	16	&	0.2	&	0.022	&	0.167	&	0.045	&	0.053	&	0.224	&	0.099	&	7.7	&	8.0	&	64.5	&	77.5	 \\
10	&	16	&	0.5	&	0.022	&	0.213	&	0.045	&	0.053	&	0.263	&	0.098	&	7.7	&	7.9	&	61.7	&	76.6	 \\
10	&	16	&	1	&	0.020	&	0.310	&	0.044	&	0.048	&	0.344	&	0.097	&	7.5	&	7.9	&	57.6	&	74.8	 \\
10	&	16	&	2	&	0.016	&	0.043	&	0.039	&	0.041	&	0.088	&	0.089	&	7.5	&	7.7	&	78.1	&	71.0	 \\
\hline																									
10	&	20	&	0	&	0.028	&	0.082	&	0.046	&	0.064	&	0.137	&	0.095	&	8.1	&	8.2	&	76.3	&	82.6	 \\
10	&	20	&	0.2	&	0.028	&	0.083	&	0.046	&	0.064	&	0.139	&	0.095	&	8.1	&	8.2	&	75.9	&	82.3	 \\
10	&	20	&	0.5	&	0.028	&	0.091	&	0.046	&	0.064	&	0.150	&	0.098	&	8.1	&	8.2	&	74.3	&	82.1	 \\
10	&	20	&	1	&	0.026	&	0.101	&	0.045	&	0.060	&	0.158	&	0.095	&	8.0	&	8.1	&	71.4	&	80.2	 \\
10	&	20	&	2	&	0.022	&	0.052	&	0.041	&	0.052	&	0.104	&	0.090	&	7.9	&	8.0	&	77.0	&	77.3	 \\
\hline																									
10	&	30	&	0	&	0.036	&	0.058	&	0.046	&	0.076	&	0.109	&	0.095	&	8.5	&	8.5	&	86.5	&	88.6	 \\
10	&	30	&	0.2	&	0.036	&	0.058	&	0.047	&	0.077	&	0.111	&	0.096	&	8.5	&	8.5	&	86.4	&	88.3	 \\
10	&	30	&	0.5	&	0.035	&	0.059	&	0.046	&	0.077	&	0.113	&	0.096	&	8.5	&	8.5	&	85.6	&	88.3	 \\
10	&	30	&	1	&	0.034	&	0.062	&	0.046	&	0.074	&	0.115	&	0.096	&	8.4	&	8.5	&	84.0	&	87.1	 \\
10	&	30	&	2	&	0.030	&	0.052	&	0.043	&	0.068	&	0.103	&	0.092	&	8.3	&	8.3	&	83.6	&	84.5	 \\
\hline																									
10	&	40	&	0	&	0.039	&	0.053	&	0.046	&	0.082	&	0.104	&	0.095	&	8.6	&	8.6	&	90.4	&	90.9	 \\
10	&	40	&	0.2	&	0.040	&	0.054	&	0.047	&	0.084	&	0.106	&	0.098	&	8.6	&	8.7	&	90.3	&	91.5	 \\
10	&	40	&	0.5	&	0.041	&	0.056	&	0.049	&	0.084	&	0.108	&	0.099	&	8.6	&	8.7	&	89.9	&	91.9	 \\
10	&	40	&	1	&	0.037	&	0.054	&	0.046	&	0.080	&	0.107	&	0.096	&	8.6	&	8.6	&	88.7	&	90.2	 \\
10	&	40	&	2	&	0.035	&	0.052	&	0.045	&	0.076	&	0.102	&	0.094	&	8.5	&	8.5	&	87.7	&	88.3	 \\
\hline																									
10	&	100	&	0	&	0.045	&	0.050	&	0.048	&	0.094	&	0.101	&	0.099	&	8.9	&	8.9	&	96.0	&	96.1	 \\
10	&	100	&	0.2	&	0.045	&	0.050	&	0.048	&	0.091	&	0.098	&	0.096	&	8.9	&	8.9	&	96.0	&	95.8	 \\
10	&	100	&	0.5	&	0.046	&	0.051	&	0.049	&	0.094	&	0.101	&	0.099	&	8.9	&	8.9	&	95.8	&	96.2	 \\
10	&	100	&	1	&	0.046	&	0.051	&	0.049	&	0.093	&	0.102	&	0.099	&	8.8	&	8.8	&	95.5	&	95.7	 \\
10	&	100	&	2	&	0.043	&	0.050	&	0.048	&	0.090	&	0.100	&	0.098	&	8.8	&	8.8	&	94.8	&	95.0	 \\
\hline																									
10	&	200	&	0	&	0.047	&	0.049	&	0.048	&	0.095	&	0.099	&	0.098	&	8.9	&	8.9	&	97.6	&	96.9	 \\
10	&	200	&	0.2	&	0.049	&	0.051	&	0.050	&	0.098	&	0.101	&	0.100	&	8.9	&	8.9	&	97.5	&	97.5	 \\
10	&	200	&	0.5	&	0.048	&	0.051	&	0.050	&	0.098	&	0.101	&	0.100	&	8.9	&	8.9	&	97.5	&	97.7	 \\
10	&	200	&	1	&	0.048	&	0.050	&	0.049	&	0.097	&	0.101	&	0.100	&	8.9	&	8.9	&	97.3	&	97.6	 \\
10	&	200	&	2	&	0.045	&	0.049	&	0.047	&	0.093	&	0.098	&	0.097	&	8.9	&	8.9	&	96.9	&	96.4	 \\
\hline																									
20	&	10	&	0	&	0.003	&	NA	&	0.103	&	0.011	&	NA	&	0.206	&	12.0	&	14.8	&	-34.5	&	 241.8	\\
20	&	10	&	0.2	&	0.003	&	NA	&	0.102	&	0.011	&	NA	&	0.205	&	11.9	&	14.8	&	-49.7	&	 240.3	\\
20	&	10	&	0.5	&	0.003	&	NA	&	0.109	&	0.010	&	NA	&	0.219	&	11.5	&	14.7	&	-105.9	&	 237.6	\\
20	&	10	&	1	&	0.002	&	NA	&	0.116	&	0.008	&	NA	&	0.231	&	11.3	&	14.5	&	-59.6	&	 231.0	\\
20	&	10	&	2	&	0.001	&	NA	&	0.031	&	0.005	&	NA	&	0.076	&	15.2	&	14.0	&	1481.0	&	 214.1	\\
\hline	
20	&	14	&	0	&	0.013	&	NA	&	0.052	&	0.033	&	NA	&	0.110	&	15.5	&	16.4	&	229.0	&	 294.8	\\
20	&	14	&	0.2	&	0.012	&	NA	&	0.051	&	0.032	&	NA	&	0.111	&	15.5	&	16.3	&	226.1	&	 294.1	\\
20	&	14	&	0.5	&	0.012	&	NA	&	0.053	&	0.031	&	NA	&	0.112	&	15.3	&	16.3	&	213.7	&	 292.4	\\
20	&	14	&	1	&	0.010	&	NA	&	0.053	&	0.027	&	NA	&	0.115	&	15.0	&	16.1	&	200.3	&	 285.7	\\
20	&	14	&	2	&	0.006	&	NA	&	0.037	&	0.020	&	NA	&	0.087	&	15.5	&	15.7	&	328.6	&	 269.9	\\
\hline																									
20	&	16	&	0	&	0.016	&	NA	&	0.048	&	0.040	&	NA	&	0.103	&	16.2	&	16.8	&	267.9	&	 311.0	\\
20	&	16	&	0.2	&	0.017	&	NA	&	0.048	&	0.040	&	NA	&	0.103	&	16.2	&	16.8	&	266.0	&	 310.6	\\
20	&	16	&	0.5	&	0.015	&	NA	&	0.048	&	0.038	&	NA	&	0.105	&	16.1	&	16.7	&	257.5	&	 307.4	\\
20	&	16	&	1	&	0.014	&	NA	&	0.049	&	0.035	&	NA	&	0.106	&	15.8	&	16.6	&	245.1	&	 302.6	\\
20	&	16	&	2	&	0.010	&	0.037	&	0.040	&	0.027	&	0.081	&	0.089	&	15.9	&	16.2	&	304.3	 &	287.1	\\
\hline																									
20	&	20	&	0	&	0.022	&	0.101	&	0.046	&	0.052	&	0.162	&	0.098	&	17.0	&	17.3	&	308.9	 &	330.7	\\
20	&	20	&	0.2	&	0.022	&	0.102	&	0.046	&	0.051	&	0.162	&	0.097	&	17.0	&	17.3	&	307.8	 &	329.2	\\
20	&	20	&	0.5	&	0.021	&	0.115	&	0.046	&	0.050	&	0.176	&	0.098	&	16.9	&	17.3	&	302.8	 &	328.4	\\
20	&	20	&	1	&	0.020	&	0.135	&	0.047	&	0.048	&	0.196	&	0.099	&	16.7	&	17.1	&	293.2	 &	323.5	\\
20	&	20	&	2	&	0.016	&	0.053	&	0.042	&	0.040	&	0.107	&	0.091	&	16.6	&	16.8	&	308.9	 &	312.0	\\
\hline																									
20	&	30	&	0	&	0.031	&	0.062	&	0.046	&	0.069	&	0.115	&	0.097	&	17.9	&	17.9	&	348.4	 &	355.0	\\
20	&	30	&	0.2	&	0.032	&	0.063	&	0.047	&	0.069	&	0.116	&	0.097	&	17.9	&	17.9	&	347.9	 &	355.9	\\
20	&	30	&	0.5	&	0.030	&	0.063	&	0.046	&	0.066	&	0.116	&	0.095	&	17.8	&	17.9	&	345.5	 &	353.3	\\
20	&	30	&	1	&	0.030	&	0.068	&	0.048	&	0.066	&	0.124	&	0.098	&	17.7	&	17.8	&	339.8	 &	351.4	\\
20	&	30	&	2	&	0.025	&	0.054	&	0.043	&	0.057	&	0.106	&	0.091	&	17.5	&	17.6	&	337.4	 &	340.8	\\
\hline																									
20	&	40	&	0	&	0.036	&	0.056	&	0.047	&	0.077	&	0.108	&	0.098	&	18.2	&	18.2	&	363.8	 &	367.3	\\
20	&	40	&	0.2	&	0.036	&	0.056	&	0.048	&	0.077	&	0.106	&	0.096	&	18.2	&	18.2	&	363.5	 &	367.3	\\
20	&	40	&	0.5	&	0.035	&	0.056	&	0.047	&	0.076	&	0.108	&	0.096	&	18.2	&	18.2	&	361.9	 &	364.8	\\
20	&	40	&	1	&	0.034	&	0.058	&	0.048	&	0.074	&	0.111	&	0.098	&	18.1	&	18.2	&	357.9	 &	364.0	\\
20	&	40	&	2	&	0.031	&	0.053	&	0.045	&	0.068	&	0.104	&	0.094	&	17.9	&	18.0	&	353.7	 &	356.3	\\
\hline
20	&	100	&	0	&	0.045	&	0.051	&	0.049	&	0.092	&	0.103	&	0.100	&	18.7	&	18.7	&	386.5	 &	388.1	\\
20	&	100	&	0.2	&	0.044	&	0.050	&	0.048	&	0.091	&	0.102	&	0.099	&	18.7	&	18.7	&	386.4	 &	387.0	\\
20	&	100	&	0.5	&	0.044	&	0.051	&	0.048	&	0.090	&	0.101	&	0.098	&	18.7	&	18.7	&	386.0	 &	385.6	\\
20	&	100	&	1	&	0.044	&	0.051	&	0.049	&	0.089	&	0.101	&	0.097	&	18.7	&	18.7	&	384.5	 &	384.8	\\
20	&	100	&	2	&	0.042	&	0.051	&	0.048	&	0.088	&	0.101	&	0.098	&	18.6	&	18.6	&	381.9	 &	382.5	\\
\hline																									
20	&	200	&	0	&	0.048	&	0.051	&	0.050	&	0.096	&	0.101	&	0.100	&	18.9	&	18.9	&	393.0	 &	394.3	\\
20	&	200	&	0.2	&	0.047	&	0.050	&	0.049	&	0.097	&	0.102	&	0.101	&	18.9	&	18.9	&	393.0	 &	393.6	\\
20	&	200	&	0.5	&	0.046	&	0.050	&	0.049	&	0.096	&	0.101	&	0.100	&	18.9	&	18.9	&	392.7	 &	393.0	\\
20	&	200	&	1	&	0.047	&	0.051	&	0.050	&	0.095	&	0.100	&	0.099	&	18.8	&	18.9	&	392.1	 &	392.8	\\
20	&	200	&	2	&	0.046	&	0.050	&	0.049	&	0.094	&	0.100	&	0.099	&	18.8	&	18.8	&	390.6	 &	391.6	\\
\hline
50	&	20	&	0	&	0.014	&	0.169	&	0.049	&	0.034	&	0.231	&	0.103	&	43.8	&	44.6	&	1945.7	 &	2064.6	\\
50	&	20	&	0.2	&	0.014	&	0.173	&	0.050	&	0.034	&	0.235	&	0.102	&	43.7	&	44.5	&	1940.1	 &	2059.9	\\
50	&	20	&	0.5	&	0.013	&	0.210	&	0.051	&	0.034	&	0.269	&	0.107	&	43.5	&	44.5	&	1914.7	 &	2057.5	\\
50	&	20	&	1	&	0.011	&	0.285	&	0.050	&	0.030	&	0.333	&	0.107	&	43.0	&	44.1	&	1862.5	 &	2022.4	\\
50	&	20	&	2	&	0.008	&	0.082	&	0.042	&	0.022	&	0.140	&	0.091	&	42.9	&	43.3	&	1907.4	 &	1950.6	\\
\hline																									
50	&	30	&	0	&	0.023	&	0.070	&	0.046	&	0.053	&	0.127	&	0.096	&	46.0	&	46.2	&	2182.5	 &	2218.9	\\
50	&	30	&	0.2	&	0.024	&	0.073	&	0.048	&	0.055	&	0.130	&	0.099	&	45.9	&	46.3	&	2179.8	 &	2226.4	\\
50	&	30	&	0.5	&	0.023	&	0.076	&	0.048	&	0.054	&	0.133	&	0.099	&	45.8	&	46.1	&	2166.7	 &	2215.6	\\
50	&	30	&	1	&	0.022	&	0.084	&	0.049	&	0.051	&	0.141	&	0.100	&	45.5	&	46.0	&	2134.5	 &	2197.4	\\
50	&	30	&	2	&	0.017	&	0.065	&	0.043	&	0.042	&	0.119	&	0.092	&	45.2	&	45.4	&	2110.0	 &	2137.9	\\
\hline																									
50	&	40	&	0	&	0.030	&	0.061	&	0.048	&	0.066	&	0.115	&	0.099	&	46.9	&	47.0	&	2277.4	 &	2298.3	\\
50	&	40	&	0.2	&	0.030	&	0.060	&	0.047	&	0.066	&	0.114	&	0.098	&	46.9	&	47.0	&	2275.6	 &	2295.2	\\
50	&	40	&	0.5	&	0.029	&	0.060	&	0.046	&	0.063	&	0.114	&	0.097	&	46.8	&	46.9	&	2266.9	 &	2290.6	\\
50	&	40	&	1	&	0.029	&	0.065	&	0.049	&	0.063	&	0.120	&	0.099	&	46.6	&	46.8	&	2243.9	 &	2277.0	\\
50	&	40	&	2	&	0.024	&	0.058	&	0.045	&	0.055	&	0.110	&	0.093	&	46.2	&	46.4	&	2214.5	 &	2234.3	\\
\hline																									
50	&	100	&	0	&	0.041	&	0.051	&	0.049	&	0.085	&	0.101	&	0.098	&	48.2	&	48.2	&	2419.8	 &	2420.3	\\
50	&	100	&	0.2	&	0.041	&	0.051	&	0.049	&	0.086	&	0.103	&	0.100	&	48.2	&	48.2	&	2419.2	 &	2416.6	\\
50	&	100	&	0.5	&	0.041	&	0.052	&	0.049	&	0.086	&	0.103	&	0.100	&	48.2	&	48.3	&	2416.3	 &	2426.5	\\
50	&	100	&	1	&	0.040	&	0.051	&	0.048	&	0.083	&	0.101	&	0.097	&	48.1	&	48.1	&	2408.0	 &	2407.3	\\
50	&	100	&	2	&	0.038	&	0.051	&	0.047	&	0.081	&	0.102	&	0.098	&	48.0	&	48.0	&	2391.7	 &	2395.3	\\
\hline
50	&	200	&	0	&	0.046	&	0.050	&	0.049	&	0.093	&	0.100	&	0.099	&	48.6	&	48.6	&	2460.7	 &	2461.6	\\
50	&	200	&	0.2	&	0.046	&	0.051	&	0.050	&	0.094	&	0.102	&	0.100	&	48.6	&	48.6	&	2460.5	 &	2459.7	\\
50	&	200	&	0.5	&	0.046	&	0.051	&	0.050	&	0.093	&	0.100	&	0.099	&	48.6	&	48.6	&	2459.1	 &	2459.3	\\
50	&	200	&	1	&	0.046	&	0.051	&	0.050	&	0.092	&	0.101	&	0.099	&	48.6	&	48.6	&	2455.1	 &	2456.8	\\
50	&	200	&	2	&	0.044	&	0.051	&	0.049	&	0.091	&	0.101	&	0.099	&	48.5	&	48.5	&	2446.5	 &	2450.8	\\
\hline
\caption{\label{tab:null2} \emph{Type I error of the standard  $Q$ test and the improved $Q$ test for homogeneity (gamma- and $\chi^2_{\e(Q)}$ approximations) under the null and moments of the distribution of $Q$.  Sample sizes are equal and balanced.  The column headings are defined in Section~\ref{subsec:simnullequal}. Here $\hat{\delta}=\sum w_i\delta_i/W$.}}

\end{longtable}
\end{footnotesize}

\newpage
      \begin{footnotesize}
   \begin{longtable} {|c|c|c||r|r|r||r|r|r||r|r|r|r|r|r|}
     \hline
$I$ &   $N$ &   $\delta$    &   $\chi^2_{.05}$  &   $\Gamma_{.05}^{th}$ &   $\Gamma_{.05}^{s}$  &    $\chi^2_{.1}$  &    $\Gamma_{.1}^{th}$ &   $\Gamma_{.1}^{s}$   &   $E_f(Q)$  &   $\bar{Q}$   &   $E_f(Q^2)$     &  $\bar{Q^2}$ &   $\var_f(Q)$   &    $s^2(Q)$   \\
\hline
5   &   20  &   0.0 &   0.035   &   0.068   &   0.070   &   0.077   &   0.120   &   0.122   &   3.7 &   3.7  &  18.5    &   20.1     &  5.2 &   6.6 \\
5   &   20  &   0.2 &   0.034   &   0.068   &   0.069   &   0.074   &   0.120   &   0.122   &   3.6 &   3.7  &  18.4    &   20.0     &  5.1 &   6.6 \\
5   &   20  &   0.5 &   0.034   &   0.074   &   0.075   &   0.075   &   0.128   &   0.129   &   3.6 &   3.7  &  17.9    &   20.1     &  4.8 &   6.6 \\
5   &   20  &   1.0 &   0.032   &   0.083   &   0.082   &   0.071   &   0.137   &   0.136   &   3.6 &   3.6  &  16.9    &   19.6     &  4.2 &   6.4 \\
5   &   20  &   2.0 &   0.028   &   0.054   &   0.050   &   0.066   &   0.108   &   0.104   &   3.5 &   3.6  &  18.1    &   18.9     &  5.8 &   6.1 \\
\hline
5   &   30  &   0.0 &   0.041   &   0.055   &   0.056   &   0.086   &   0.107   &   0.107   &   3.8 &   3.8  &  21.1    &   21.8     &  6.7 &   7.2 \\
5   &   30  &   0.2 &   0.041   &   0.056   &   0.056   &   0.086   &   0.107   &   0.107   &   3.8 &   3.8  &  21.0    &   21.7     &  6.6 &   7.3 \\
5   &   30  &   0.5 &   0.039   &   0.056   &   0.056   &   0.084   &   0.107   &   0.107   &   3.8 &   3.8  &  20.8    &   21.5     &  6.5 &   7.1 \\
5   &   30  &   1.0 &   0.039   &   0.058   &   0.059   &   0.083   &   0.111   &   0.111   &   3.8 &   3.8  &  20.3    &   21.3     &  6.2 &   7.1 \\
5   &   30  &   2.0 &   0.036   &   0.054   &   0.053   &   0.078   &   0.106   &   0.106   &   3.7 &   3.7  &  20.1    &   20.7     &  6.4 &   6.8 \\
\hline
5   &   40  &   0.0 &   0.043   &   0.053   &   0.053   &   0.089   &   0.103   &   0.104   &   3.9 &   3.9  &  22.0    &   22.3     &  7.1 &   7.4 \\
5   &   40  &   0.2 &   0.043   &   0.053   &   0.053   &   0.089   &   0.103   &   0.103   &   3.9 &   3.9  &  22.0    &   22.4     &  7.1 &   7.4 \\
5   &   40  &   0.5 &   0.043   &   0.054   &   0.054   &   0.089   &   0.103   &   0.103   &   3.9 &   3.9  &  21.9    &   22.3     &  7.0 &   7.4 \\
5   &   40  &   1.0 &   0.042   &   0.054   &   0.054   &   0.089   &   0.106   &   0.105   &   3.8 &   3.8  &  21.5    &   22.1     &  6.9 &   7.3 \\
5   &   40  &   2.0 &   0.039   &   0.052   &   0.049   &   0.084   &   0.104   &   0.102   &   3.8 &   3.8  &  21.2    &   21.5     &  6.9 &   7.0 \\
\hline
5   &   100 &   0.0 &   0.047   &   0.050   &   0.050   &   0.095   &   0.100   &   0.100   &   4.0 &   4.0  &  23.3    &   23.3     &  7.7 &   7.8 \\
5   &   100 &   0.2 &   0.047   &   0.050   &   0.050   &   0.096   &   0.100   &   0.100   &   4.0 &   4.0  &  23.3    &   23.4     &  7.7 &   7.8 \\
5   &   100 &   0.5 &   0.048   &   0.052   &   0.052   &   0.097   &   0.102   &   0.102   &   4.0 &   4.0  &  23.3    &   23.5     &  7.7 &   7.9 \\
5   &   100 &   1.0 &   0.047   &   0.050   &   0.050   &   0.096   &   0.102   &   0.102   &   3.9 &   4.0  &  23.2    &   23.4     &  7.7 &   7.8 \\
5   &   100 &   2.0 &   0.045   &   0.049   &   0.049   &   0.093   &   0.100   &   0.100   &   3.9 &   3.9  &  23.0    &   23.0     &  7.6 &   7.6 \\
\hline
5   &   200 &   0.0 &   0.048   &   0.050   &   0.050   &   0.097   &   0.099   &   0.099   &   4.0 &   4.0  &  23.7    &   23.5     &  7.9 &   7.8 \\
5   &   200 &   0.2 &   0.049   &   0.050   &   0.050   &   0.098   &   0.100   &   0.100   &   4.0 &   4.0  &  23.7    &   23.8     &  7.9 &   7.9 \\
5   &   200 &   0.5 &   0.049   &   0.050   &   0.050   &   0.099   &   0.101   &   0.101   &   4.0 &   4.0  &  23.7    &   23.7     &  7.9 &   7.9 \\
5   &   200 &   1.0 &   0.049   &   0.050   &   0.050   &   0.098   &   0.100   &   0.100   &   4.0 &   4.0  &  23.6    &   23.7     &  7.9 &   7.9 \\
5   &   200 &   2.0 &   0.048   &   0.050   &   0.050   &   0.097   &   0.101   &   0.101   &   4.0 &   4.0  &  23.5    &   23.4     &  7.8 &   7.8 \\
\hline
10  &   20  &   0.0 &   0.030   &   0.082   &   0.083   &   0.065   &   0.137   &   0.139   &   8.1 &   8.2  &  76.3    &   82.7     &  10.3    &   14.9    \\
10  &   20  &   0.2 &   0.027   &   0.081   &   0.082   &   0.063   &   0.138   &   0.139   &   8.1 &   8.2  &  75.9    &   82.1     &  10.1    &   14.7    \\
10  &   20  &   0.5 &   0.028   &   0.090   &   0.091   &   0.061   &   0.147   &   0.148   &   8.1 &   8.2  &  74.3    &   81.7     &  9.2 &   14.5    \\
10  &   20  &   1.0 &   0.025   &   0.104   &   0.102   &   0.060   &   0.163   &   0.161   &   8.0 &   8.1  &  71.4    &   80.4     &  8.0 &   14.3    \\
10  &   20  &   2.0 &   0.022   &   0.043   &   0.039   &   0.051   &   0.093   &   0.089   &   7.9 &   8.0  &  77.0    &   77.0     &  15.1    &   13.5    \\
\hline
10  &   30  &   0.0 &   0.037   &   0.059   &   0.059   &   0.078   &   0.111   &   0.111   &   8.5 &   8.5  &  86.5    &   88.8     &  14.4    &   16.1    \\
10  &   30  &   0.2 &   0.036   &   0.059   &   0.059   &   0.078   &   0.111   &   0.111   &   8.5 &   8.5  &  86.4    &   88.6     &  14.4    &   16.0    \\
10  &   30  &   0.5 &   0.035   &   0.059   &   0.060   &   0.077   &   0.113   &   0.113   &   8.5 &   8.5  &  85.6    &   88.1     &  14.1    &   15.9    \\
10  &   30  &   1.0 &   0.034   &   0.062   &   0.062   &   0.073   &   0.114   &   0.114   &   8.4 &   8.4  &  84.0    &   86.8     &  13.5    &   15.6    \\
10  &   30  &   2.0 &   0.031   &   0.053   &   0.053   &   0.070   &   0.106   &   0.105   &   8.3 &   8.4  &  83.6    &   85.0     &  14.7    &   15.2    \\
\hline
10  &   40  &   0.0 &   0.040   &   0.054   &   0.054   &   0.084   &   0.105   &   0.106   &   8.6 &   8.7  &  90.4    &   91.6     &  15.7    &   16.6    \\
10  &   40  &   0.2 &   0.040   &   0.054   &   0.055   &   0.084   &   0.105   &   0.106   &   8.6 &   8.7  &  90.3    &   91.5     &  15.7    &   16.7    \\
10  &   40  &   0.5 &   0.040   &   0.055   &   0.055   &   0.082   &   0.104   &   0.105   &   8.6 &   8.6  &  89.9    &   91.1     &  15.5    &   16.5    \\
10  &   40  &   1.0 &   0.038   &   0.055   &   0.054   &   0.080   &   0.107   &   0.105   &   8.6 &   8.6  &  88.7    &   90.3     &  15.2    &   16.2    \\
10  &   40  &   2.0 &   0.034   &   0.050   &   0.048   &   0.075   &   0.101   &   0.099   &   8.5 &   8.5  &  87.7    &   88.1     &  15.5    &   15.7    \\
\hline
10  &   100 &   0.0 &   0.047   &   0.051   &   0.051   &   0.095   &   0.103   &   0.103   &   8.9 &   8.9  &  96.0    &   96.5     &  17.3    &   17.7    \\
10  &   100 &   0.2 &   0.047   &   0.051   &   0.051   &   0.094   &   0.101   &   0.101   &   8.9 &   8.9  &  96.0    &   96.4     &  17.3    &   17.5    \\
10  &   100 &   0.5 &   0.046   &   0.051   &   0.051   &   0.093   &   0.100   &   0.100   &   8.9 &   8.9  &  95.8    &   96.0     &  17.3    &   17.4    \\
10  &   100 &   1.0 &   0.044   &   0.050   &   0.050   &   0.091   &   0.099   &   0.099   &   8.9 &   8.8  &  95.5    &   95.1     &  17.2    &   17.2    \\
10  &   100 &   2.0 &   0.044   &   0.050   &   0.050   &   0.090   &   0.099   &   0.099   &   8.8 &   8.8  &  94.8    &   94.5     &  17.1    &   17.1    \\
\hline
10  &   200 &   0.0 &   0.047   &   0.050   &   0.050   &   0.097   &   0.101   &   0.101   &   8.9 &   8.9  &  97.6    &   97.4     &  17.7    &   17.6    \\
10  &   200 &   0.2 &   0.049   &   0.051   &   0.051   &   0.098   &   0.101   &   0.101   &   8.9 &   9.0  &  97.6    &   98.0     &  17.7    &   17.8    \\
10  &   200 &   0.5 &   0.048   &   0.050   &   0.050   &   0.097   &   0.100   &   0.100   &   8.9 &   8.9  &  97.5    &   97.6     &  17.7    &   17.7    \\
10  &   200 &   1.0 &   0.047   &   0.050   &   0.050   &   0.097   &   0.101   &   0.101   &   8.9 &   9.0  &  97.3    &   97.8     &  17.6    &   17.7    \\
10  &   200 &   2.0 &   0.046   &   0.049   &   0.049   &   0.094   &   0.099   &   0.099   &   8.9 &   8.9  &  96.9    &   96.9     &  17.6    &   17.4    \\
\hline
20  &   20  &   0.0 &   0.022   &   0.100   &   0.101   &   0.053   &   0.160   &   0.161   &   17.0    &    17.3   &   308.9    &  330.5   &   18.6    &   30.9    \\
20  &   20  &   0.2 &   0.022   &   0.102   &   0.103   &   0.052   &   0.161   &   0.162   &   17.0    &    17.3   &   307.8    &  329.0   &   18.2    &   31.0    \\
20  &   20  &   0.5 &   0.022   &   0.117   &   0.118   &   0.050   &   0.179   &   0.179   &   16.9    &    17.3   &   302.8    &  328.6   &   16.0    &   30.8    \\
20  &   20  &   1.0 &   0.019   &   0.138   &   0.136   &   0.047   &   0.201   &   0.199   &   16.7    &    17.1   &   293.2    &  323.1   &   13.2    &   30.0    \\
20  &   20  &   2.0 &   0.015   &   0.042   &   0.038   &   0.039   &   0.092   &   0.087   &   16.6    &    16.8   &   308.9    &  310.7   &   32.5    &   28.2    \\
\hline
20  &   30  &   0.0 &   0.032   &   0.063   &   0.063   &   0.070   &   0.117   &   0.117   &   17.9    &    18.0   &   348.4    &  357.5   &   29.3    &   33.8    \\
20  &   30  &   0.2 &   0.032   &   0.063   &   0.063   &   0.069   &   0.117   &   0.118   &   17.9    &    18.0   &   347.9    &  356.4   &   29.2    &   33.8    \\
20  &   30  &   0.5 &   0.031   &   0.063   &   0.064   &   0.067   &   0.119   &   0.119   &   17.8    &    17.9   &   345.5    &  354.2   &   28.4    &   33.4    \\
20  &   30  &   1.0 &   0.029   &   0.066   &   0.066   &   0.064   &   0.121   &   0.121   &   17.7    &    17.8   &   339.8    &  349.0   &   27.1    &   32.7    \\
20  &   30  &   2.0 &   0.025   &   0.053   &   0.052   &   0.058   &   0.105   &   0.104   &   17.5    &    17.6   &   337.4    &  340.8   &   30.7    &   31.6    \\
\hline
20  &   40  &   0.0 &   0.035   &   0.054   &   0.055   &   0.076   &   0.107   &   0.108   &   18.2    &    18.2   &   363.8    &  366.8   &   32.6    &   34.7    \\
20  &   40  &   0.2 &   0.035   &   0.054   &   0.055   &   0.076   &   0.107   &   0.107   &   18.2    &    18.2   &   363.5    &  366.8   &   32.5    &   34.5    \\
20  &   40  &   0.5 &   0.035   &   0.056   &   0.056   &   0.076   &   0.109   &   0.109   &   18.2    &    18.2   &   361.9    &  365.7   &   32.1    &   34.7    \\
20  &   40  &   1.0 &   0.035   &   0.058   &   0.056   &   0.073   &   0.110   &   0.109   &   18.1    &    18.2   &   357.9    &  364.0   &   31.3    &   34.3    \\
20  &   40  &   2.0 &   0.030   &   0.052   &   0.050   &   0.068   &   0.102   &   0.100   &   17.9    &    17.9   &   353.7    &  355.1   &   32.4    &   33.1    \\
\hline
20  &   100 &   0.0 &   0.044   &   0.051   &   0.051   &   0.091   &   0.101   &   0.101   &   18.7    &    18.7   &   386.5    &  386.7   &   36.4    &   36.9    \\
20  &   100 &   0.2 &   0.045   &   0.052   &   0.052   &   0.092   &   0.102   &   0.102   &   18.7    &    18.7   &   386.4    &  387.5   &   36.4    &   37.0    \\
20  &   100 &   0.5 &   0.044   &   0.051   &   0.051   &   0.091   &   0.102   &   0.102   &   18.7    &    18.7   &   386.0    &  386.7   &   36.3    &   36.7    \\
20  &   100 &   1.0 &   0.044   &   0.051   &   0.051   &   0.089   &   0.101   &   0.101   &   18.7    &    18.7   &   384.5    &  385.2   &   36.1    &   36.6    \\
20  &   100 &   2.0 &   0.042   &   0.049   &   0.049   &   0.086   &   0.100   &   0.100   &   18.6    &    18.6   &   381.9    &  382.4   &   36.0    &   36.0    \\
\hline
20  &   200 &   0.0 &   0.047   &   0.050   &   0.050   &   0.095   &   0.100   &   0.100   &   18.9    &    18.8   &   393.0    &  391.9   &   37.3    &   37.5    \\
20  &   200 &   0.2 &   0.046   &   0.049   &   0.049   &   0.094   &   0.099   &   0.099   &   18.9    &    18.9   &   393.0    &  392.9   &   37.3    &   37.1    \\
20  &   200 &   0.5 &   0.047   &   0.050   &   0.050   &   0.096   &   0.101   &   0.101   &   18.9    &    18.9   &   392.7    &  394.7   &   37.3    &   37.6    \\
20  &   200 &   1.0 &   0.047   &   0.051   &   0.051   &   0.094   &   0.099   &   0.099   &   18.8    &    18.8   &   392.1    &  391.0   &   37.2    &   37.3    \\
20  &   200 &   2.0 &   0.045   &   0.049   &   0.049   &   0.092   &   0.099   &   0.099   &   18.8    &    18.8   &   390.6    &  389.9   &   37.1    &   36.7    \\
\hline
50  &   20  &   0   &   0.014   &   0.169   &   0.169   &   0.034   &   0.230   &   0.231   &   43.8    &    44.6   &   1945.7  &   2064.8  &   29.2    &   79.4    \\
50  &   20  &   0.2 &   0.014   &   0.176   &   0.177   &   0.035   &   0.239   &   0.240   &   43.7    &    44.6   &   1940.1  &   2065.0  &   27.6    &   80.0    \\
50  &   20  &   0.5 &   0.014   &   0.214   &   0.216   &   0.033   &   0.274   &   0.275   &   43.5    &    44.4   &   1914.7  &   2054.5  &   20.6    &   79.0    \\
50  &   20  &   1   &   0.012   &   0.290   &   0.288   &   0.030   &   0.338   &   0.336   &   43.0    &    44.1   &   1862.5  &   2019.2  &   10.5    &   76.8    \\
50  &   20  &   2   &   0.008   &   0.061   &   0.057   &   0.022   &   0.116   &   0.112   &   42.9    &    43.3   &   1907.4  &   1949.7  &   67.4    &   72.8    \\
\hline
50  &   30  &   0   &   0.025   &   0.073   &   0.073   &   0.055   &   0.130   &   0.130   &   46.0    &    46.3   &   2182.5   &  2226.6  &       69.2&87.0       \\
50  &   30  &   0.2 &   0.024   &   0.073   &   0.073   &   0.054   &   0.129   &   0.129   &   45.9    &    46.2   &   2179.8   &  2224.8  &   68.7& 87.0      \\
50  &   30  &   0.5 &   0.023   &   0.076   &   0.076   &   0.052   &   0.133   &   0.133   &   45.8    &    46.1   &   2166.7   &  2213.2  &   66.3 & 85.9         \\
50  &   30  &   1   &   0.022   &   0.083   &   0.083   &   0.050   &   0.142   &   0.142   &   45.5    &    45.9   &   2134.5   &  2195.9  &   61.8& 84.7          \\
50  &   30  &   2   &   0.017   &   0.063   &   0.062   &   0.042   &   0.116   &   0.115   &   45.2    &    45.3   &   2110.0   &  2136.7  &   71.4 &80.6      \\
\hline
50  &   40  &   0.0 &   0.030   &   0.060   &   0.060   &   0.066   &   0.113   &   0.113   &   46.9    &    47.0   &   2277.4   &  2299.9  &   80.8    &   89.6    \\
50  &   40  &   0.2 &   0.029   &   0.059   &   0.059   &   0.065   &   0.113   &   0.113   &   46.9    &    47.0   &   2275.6   &  2297.9  &   80.6    &   89.2    \\
50  &   40  &   0.5 &   0.030   &   0.062   &   0.062   &   0.065   &   0.114   &   0.115   &   46.8    &    46.9   &   2266.9   &  2290.7  &   79.3    &   89.6    \\
50  &   40  &   1.0 &   0.028   &   0.065   &   0.065   &   0.062   &   0.118   &   0.119   &   46.6    &    46.8   &   2243.9   &  2277.9  &   76.7    &   88.1    \\
50  &   40  &   2.0 &   0.024   &   0.057   &   0.057   &   0.054   &   0.109   &   0.109   &   46.2    &    46.3   &   2214.5   &  2231.7  &   79.3    &   85.6    \\
\hline
50  &   100 &   0.0 &   0.041   &   0.051   &   0.052   &   0.086   &   0.102   &   0.102   &   48.2    &    48.2   &   2419.8   &  2419.3  &   93.5    &   94.7    \\
50  &   100 &   0.2 &   0.042   &   0.052   &   0.052   &   0.086   &   0.102   &   0.102   &   48.2    &    48.2   &   2419.2   &  2421.7  &   93.4    &   94.6    \\
50  &   100 &   0.5 &   0.041   &   0.052   &   0.052   &   0.085   &   0.101   &   0.101   &   48.2    &    48.2   &   2416.3   &  2419.5  &   93.2    &   94.2    \\
50  &   100 &   1.0 &   0.040   &   0.051   &   0.051   &   0.084   &   0.102   &   0.102   &   48.1    &    48.1   &   2408.0   &  2411.3  &   92.6    &   94.2    \\
50  &   100 &   2.0 &   0.038   &   0.051   &   0.051   &   0.081   &   0.101   &   0.101   &   48.0    &    48.0   &   2391.7   &  2394.1  &   92.2    &   93.1    \\
\hline
50  &   200 &   0   &   0.046   &   0.051   &   0.051   &   0.095   &   0.102   &   0.102   &   48.6    &    48.7   &   2460.7  &   2466.4  &   96.0    &   96.8    \\
50  &   200 &   0.2 &   0.045   &   0.050   &   0.050   &   0.093   &   0.101   &   0.101   &   48.6    &    48.6   &   2460.5  &   2460.2  &   96.0    &   96.6    \\
50  &   200 &   0.5 &   0.045   &   0.050   &   0.050   &   0.093   &   0.100   &   0.100   &   48.6    &    48.6   &   2459.1  &   2456.6  &   95.9    &   96.5    \\
50  &   200 &   1   &   0.046   &   0.051   &   0.051   &   0.093   &   0.102   &   0.102   &   48.6    &    48.6   &   2455.1  &   2458.3  &   95.7    &   96.2    \\
50  &   200 &   2   &   0.043   &   0.049   &   0.049   &   0.089   &   0.099   &   0.099   &   48.5    &    48.4   &   2446.5  &   2439.0  &   95.4    &   95.7    \\

\hline

\caption{\label{tab:null1} \emph{Type I error of the standard  $Q$ test and the improved $Q$ test for homogeneity under the null and moments of the distribution of $Q$.  Sample sizes are equal and balanced.  The column headings are defined in Section~\ref{subsec:simnullequal}. Here $\hat{\delta}=\sum A_i^{-1}\delta_i/\sum A_i^{-1}$.}}

\end{longtable}
\end{footnotesize}

\pagebreak
     \begin{footnotesize}
   \begin{longtable} {|c|c||r|r|r||r|r|r||r|r|r|r|r|r|}
     \hline
$I$ &   $\bar{N}$   &   $\chi^2_{.05}$  &   $\Gamma_{.05}^{s}$ &   $\chi^2_{E(Q),.05}$  &   $\chi^2_{.1}$    &   $\Gamma_{.1}^{s}$ &   $\chi^2_{E(Q),.1}$   &   $E_f(Q)$  &   $\bar{Q}$   &   $E_f(Q^2)$    &   $\bar{Q^2}$  &  $\var_f(Q)$   &    $s^2(Q)$    \\
\hline
5	&	60	&	0.041	&	0.056	&	0.048	&	0.086	&	0.107	&	0.097	&	3.8	&	3.8	&	21.1	&	21.8	&	 6.7	&	7.2	\\
5	&	100	&	0.046	&	0.051	&	0.048	&	0.095	&	0.102	&	0.099	&	3.9	&	3.9	&	23.0	&	23.2	&	 7.6	&	7.7	\\
5	&	160	&	0.050	&	0.052	&	0.051	&	0.098	&	0.101	&	0.100	&	4.0	&	4.0	&	23.5	&	23.7	&	 7.8	&	7.9	\\
\hline																											
10	&	60	&	0.038	&	0.057	&	0.047	&	0.081	&	0.110	&	0.098	&	8.6	&	8.6	&	88.0	&	90.4	&	 14.8	&	16.3	\\
10	&	100	&	0.045	&	0.051	&	0.049	&	0.093	&	0.102	&	0.099	&	8.8	&	8.9	&	95.0	&	95.7	&	 17.1	&	17.3	\\
10	&	160	&	0.048	&	0.051	&	0.049	&	0.095	&	0.100	&	0.098	&	8.9	&	8.9	&	96.9	&	96.7	&	 17.5	&	17.6	\\
\hline																											
20	&	60	&	0.034	&	0.060	&	0.048	&	0.074	&	0.113	&	0.098	&	18.0	&	18.1	&	356.3	&	 363.5	&	30.7	&	34.5	\\
20	&	100	&	0.043	&	0.051	&	0.048	&	0.088	&	0.101	&	0.097	&	18.6	&	18.6	&	382.8	&	 383.0	&	35.9	&	36.3	\\
20	&	160	&	0.046	&	0.050	&	0.049	&	0.093	&	0.100	&	0.098	&	18.8	&	18.8	&	390.3	&	 389.6	&	36.9	&	37.1	\\

\hline

\caption{\label{tab:null3} \emph{Type I error of the standard  $Q$ test and the improved $Q$ test for homogeneity under the null and moments of the distribution of $Q$.  Sample sizes are unequal but balanced.  The column headings are defined in Section~\ref{subsec:simnullequal}.}}

\end{longtable}
\end{footnotesize}

\newpage
     \begin{footnotesize}
   \begin{longtable} {|c|c|c||r|r|r||r|r|r|}
     \hline
& & & \multicolumn{3}{c||} {power at level 0.05} & \multicolumn{3}{c|} {power at level 0.10}\\
\hline
$I$ &   $N$ &   $\tau^2$    &   $\chi^2_{K-1}$  & Gamma  &  $\chi^2_{\e(Q)}$ &  $\chi^2_{K-1}$  & Gamma  &  $\chi^2_{\e(Q)}$\\
\hline
5	&	20	&	0.025	&	0.051	&	0.099	&	0.065	&	0.100	&	0.160	&	0.127	\\
5	&	20	&	0.05	&	0.071	&	0.134	&	0.089	&	0.136	&	0.203	&	0.163	\\
5	&	20	&	0.1	&	0.121	&	0.208	&	0.148	&	0.210	&	0.294	&	0.245	\\
5	&	20	&	0.15	&	0.177	&	0.274	&	0.209	&	0.277	&	0.369	&	0.316	\\
5	&	20	&	0.2	&	0.230	&	0.340	&	0.266	&	0.343	&	0.438	&	0.387	\\
5	&	20	&	0.25	&	0.281	&	0.397	&	0.321	&	0.399	&	0.500	&	0.442	\\
\hline																	
5	&	30	&	0.025	&	0.072	&	0.095	&	0.082	&	0.134	&	0.164	&	0.150	\\
5	&	30	&	0.05	&	0.111	&	0.142	&	0.125	&	0.190	&	0.228	&	0.210	\\
5	&	30	&	0.1	&	0.196	&	0.236	&	0.214	&	0.297	&	0.339	&	0.320	\\
5	&	30	&	0.15	&	0.286	&	0.336	&	0.307	&	0.402	&	0.447	&	0.426	\\
5	&	30	&	0.2	&	0.369	&	0.416	&	0.388	&	0.481	&	0.522	&	0.503	\\
5	&	30	&	0.25	&	0.444	&	0.491	&	0.463	&	0.551	&	0.591	&	0.574	\\
\hline																	
5	&	40	&	0.025	&	0.092	&	0.107	&	0.099	&	0.161	&	0.182	&	0.174	\\
5	&	40	&	0.05	&	0.147	&	0.168	&	0.156	&	0.232	&	0.258	&	0.247	\\
5	&	40	&	0.1	&	0.271	&	0.302	&	0.286	&	0.383	&	0.413	&	0.401	\\
5	&	40	&	0.15	&	0.387	&	0.418	&	0.402	&	0.500	&	0.528	&	0.515	\\
5	&	40	&	0.2	&	0.486	&	0.516	&	0.501	&	0.592	&	0.612	&	0.605	\\
5	&	40	&	0.25	&	0.560	&	0.590	&	0.575	&	0.662	&	0.683	&	0.674	\\
\hline																	
5	&	50	&	0.025	&	0.110	&	0.122	&	0.117	&	0.182	&	0.199	&	0.192	\\
5	&	50	&	0.05	&	0.196	&	0.214	&	0.204	&	0.292	&	0.313	&	0.304	\\
5	&	50	&	0.1	&	0.330	&	0.354	&	0.342	&	0.447	&	0.467	&	0.459	\\
5	&	50	&	0.15	&	0.465	&	0.488	&	0.478	&	0.572	&	0.588	&	0.582	\\
5	&	50	&	0.2	&	0.570	&	0.589	&	0.581	&	0.663	&	0.680	&	0.674	\\
5	&	50	&	0.25	&	0.650	&	0.671	&	0.661	&	0.731	&	0.746	&	0.740	\\
\hline																	
5	&	60	&	0.025	&	0.127	&	0.138	&	0.132	&	0.210	&	0.224	&	0.217	\\
5	&	60	&	0.05	&	0.225	&	0.242	&	0.233	&	0.326	&	0.342	&	0.336	\\
5	&	60	&	0.1	&	0.394	&	0.411	&	0.403	&	0.501	&	0.517	&	0.510	\\
5	&	60	&	0.15	&	0.531	&	0.547	&	0.539	&	0.633	&	0.645	&	0.641	\\
5	&	60	&	0.2	&	0.633	&	0.647	&	0.640	&	0.718	&	0.729	&	0.725	\\
5	&	60	&	0.25	&	0.707	&	0.720	&	0.714	&	0.782	&	0.791	&	0.788	\\
\hline																	
5	&	80	&	0.025	&	0.160	&	0.170	&	0.165	&	0.254	&	0.264	&	0.261	\\
5	&	80	&	0.05	&	0.295	&	0.309	&	0.303	&	0.406	&	0.417	&	0.413	\\
5	&	80	&	0.1	&	0.496	&	0.509	&	0.504	&	0.598	&	0.608	&	0.604	\\
5	&	80	&	0.15	&	0.639	&	0.648	&	0.643	&	0.717	&	0.724	&	0.721	\\
5	&	80	&	0.2	&	0.728	&	0.736	&	0.732	&	0.798	&	0.804	&	0.802	\\
5	&	80	&	0.25	&	0.800	&	0.806	&	0.803	&	0.853	&	0.857	&	0.856	\\
\hline																	
10	&	20	&	0.025	&	0.050	&	0.141	&	0.078	&	0.103	&	0.211	&	0.150	\\
10	&	20	&	0.05	&	0.083	&	0.195	&	0.118	&	0.151	&	0.277	&	0.205	\\
10	&	20	&	0.1	&	0.161	&	0.310	&	0.215	&	0.258	&	0.414	&	0.324	\\
10	&	20	&	0.15	&	0.256	&	0.432	&	0.319	&	0.371	&	0.531	&	0.445	\\
10	&	20	&	0.2	&	0.347	&	0.532	&	0.417	&	0.471	&	0.628	&	0.544	\\
10	&	20	&	0.25	&	0.445	&	0.627	&	0.515	&	0.567	&	0.708	&	0.640	\\
\hline																	
10	&	30	&	0.025	&	0.081	&	0.121	&	0.097	&	0.150	&	0.201	&	0.180	\\
10	&	30	&	0.05	&	0.147	&	0.210	&	0.178	&	0.243	&	0.305	&	0.279	\\
10	&	30	&	0.1	&	0.294	&	0.369	&	0.331	&	0.410	&	0.486	&	0.455	\\
10	&	30	&	0.15	&	0.438	&	0.520	&	0.482	&	0.561	&	0.628	&	0.600	\\
10	&	30	&	0.2	&	0.564	&	0.638	&	0.602	&	0.676	&	0.733	&	0.709	\\
10	&	30	&	0.25	&	0.668	&	0.732	&	0.702	&	0.764	&	0.811	&	0.791	\\
\hline																	
10	&	40	&	0.025	&	0.106	&	0.135	&	0.121	&	0.184	&	0.222	&	0.207	\\
10	&	40	&	0.05	&	0.203	&	0.244	&	0.223	&	0.310	&	0.354	&	0.338	\\
10	&	40	&	0.1	&	0.411	&	0.464	&	0.440	&	0.538	&	0.582	&	0.568	\\
10	&	40	&	0.15	&	0.600	&	0.642	&	0.627	&	0.703	&	0.738	&	0.725	\\
10	&	40	&	0.2	&	0.719	&	0.753	&	0.736	&	0.799	&	0.826	&	0.815	\\
10	&	40	&	0.25	&	0.803	&	0.833	&	0.821	&	0.865	&	0.885	&	0.877	\\
\hline																	
10	&	50	&	0.025	&	0.136	&	0.159	&	0.151	&	0.224	&	0.258	&	0.246	\\
10	&	50	&	0.05	&	0.267	&	0.299	&	0.285	&	0.381	&	0.415	&	0.404	\\
10	&	50	&	0.1	&	0.524	&	0.556	&	0.544	&	0.634	&	0.660	&	0.652	\\
10	&	50	&	0.15	&	0.698	&	0.730	&	0.716	&	0.783	&	0.806	&	0.798	\\
10	&	50	&	0.2	&	0.809	&	0.830	&	0.822	&	0.872	&	0.888	&	0.882	\\
10	&	50	&	0.25	&	0.878	&	0.891	&	0.886	&	0.919	&	0.928	&	0.925	\\
\hline																	
10	&	60	&	0.025	&	0.170	&	0.191	&	0.181	&	0.268	&	0.292	&	0.284	\\
10	&	60	&	0.05	&	0.338	&	0.366	&	0.354	&	0.458	&	0.486	&	0.478	\\
10	&	60	&	0.1	&	0.619	&	0.644	&	0.635	&	0.717	&	0.736	&	0.730	\\
10	&	60	&	0.15	&	0.780	&	0.797	&	0.791	&	0.845	&	0.858	&	0.853	\\
10	&	60	&	0.2	&	0.862	&	0.875	&	0.869	&	0.909	&	0.917	&	0.914	\\
10	&	60	&	0.25	&	0.917	&	0.927	&	0.923	&	0.948	&	0.953	&	0.951	\\
\hline																	
10	&	80	&	0.025	&	0.229	&	0.245	&	0.239	&	0.332	&	0.347	&	0.343	\\
10	&	80	&	0.05	&	0.459	&	0.477	&	0.469	&	0.572	&	0.591	&	0.585	\\
10	&	80	&	0.1	&	0.734	&	0.748	&	0.742	&	0.811	&	0.822	&	0.819	\\
10	&	80	&	0.15	&	0.877	&	0.887	&	0.884	&	0.919	&	0.925	&	0.923	\\
10	&	80	&	0.2	&	0.940	&	0.944	&	0.942	&	0.961	&	0.965	&	0.963	\\
10	&	80	&	0.25	&	0.962	&	0.965	&	0.964	&	0.975	&	0.977	&	0.977	\\
\hline																	
20	&	20	&	0.025	&	0.051	&	0.199	&	0.100	&	0.106	&	0.277	&	0.178	\\
20	&	20	&	0.05	&	0.096	&	0.298	&	0.160	&	0.170	&	0.391	&	0.270	\\
20	&	20	&	0.1	&	0.217	&	0.479	&	0.314	&	0.329	&	0.577	&	0.448	\\
20	&	20	&	0.15	&	0.383	&	0.659	&	0.506	&	0.520	&	0.738	&	0.632	\\
20	&	20	&	0.2	&	0.523	&	0.780	&	0.638	&	0.652	&	0.841	&	0.754	\\
20	&	20	&	0.25	&	0.655	&	0.862	&	0.751	&	0.761	&	0.901	&	0.841	\\
\hline																	
20	&	30	&	0.025	&	0.098	&	0.169	&	0.132	&	0.174	&	0.259	&	0.225	\\
20	&	30	&	0.05	&	0.197	&	0.298	&	0.256	&	0.306	&	0.413	&	0.372	\\
20	&	30	&	0.1	&	0.451	&	0.566	&	0.514	&	0.576	&	0.680	&	0.642	\\
20	&	30	&	0.15	&	0.670	&	0.765	&	0.724	&	0.770	&	0.838	&	0.815	\\
20	&	30	&	0.2	&	0.798	&	0.869	&	0.839	&	0.874	&	0.918	&	0.903	\\
20	&	30	&	0.25	&	0.888	&	0.929	&	0.913	&	0.932	&	0.957	&	0.949	\\
\hline																	
20	&	40	&	0.025	&	0.138	&	0.193	&	0.170	&	0.235	&	0.296	&	0.277	\\
20	&	40	&	0.05	&	0.311	&	0.379	&	0.353	&	0.429	&	0.500	&	0.478	\\
20	&	40	&	0.1	&	0.627	&	0.695	&	0.667	&	0.736	&	0.786	&	0.770	\\
20	&	40	&	0.15	&	0.818	&	0.858	&	0.842	&	0.885	&	0.911	&	0.903	\\
20	&	40	&	0.2	&	0.917	&	0.938	&	0.930	&	0.949	&	0.961	&	0.958	\\
20	&	40	&	0.25	&	0.962	&	0.972	&	0.968	&	0.980	&	0.986	&	0.984	\\
\hline																	
20	&	50	&	0.025	&	0.189	&	0.230	&	0.214	&	0.293	&	0.342	&	0.326	\\
20	&	50	&	0.05	&	0.407	&	0.460	&	0.441	&	0.534	&	0.586	&	0.573	\\
20	&	50	&	0.1	&	0.758	&	0.797	&	0.781	&	0.841	&	0.867	&	0.859	\\
20	&	50	&	0.15	&	0.913	&	0.928	&	0.924	&	0.947	&	0.960	&	0.956	\\
20	&	50	&	0.2	&	0.968	&	0.976	&	0.973	&	0.983	&	0.988	&	0.986	\\
20	&	50	&	0.25	&	0.987	&	0.990	&	0.989	&	0.993	&	0.995	&	0.994	\\
\hline																	
20	&	60	&	0.025	&	0.241	&	0.278	&	0.266	&	0.356	&	0.392	&	0.382	\\
20	&	60	&	0.05	&	0.502	&	0.547	&	0.531	&	0.627	&	0.664	&	0.654	\\
20	&	60	&	0.1	&	0.843	&	0.865	&	0.856	&	0.901	&	0.914	&	0.911	\\
20	&	60	&	0.15	&	0.951	&	0.959	&	0.955	&	0.971	&	0.975	&	0.974	\\
20	&	60	&	0.2	&	0.985	&	0.989	&	0.988	&	0.993	&	0.994	&	0.994	\\
20	&	60	&	0.25	&	0.995	&	0.996	&	0.996	&	0.998	&	0.998	&	0.998	\\
\hline																	
20	&	80	&	0.025	&	0.341	&	0.369	&	0.360	&	0.467	&	0.497	&	0.488	\\
20	&	80	&	0.05	&	0.671	&	0.697	&	0.687	&	0.768	&	0.787	&	0.782	\\
20	&	80	&	0.1	&	0.935	&	0.943	&	0.942	&	0.962	&	0.966	&	0.966	\\
20	&	80	&	0.15	&	0.985	&	0.988	&	0.987	&	0.992	&	0.993	&	0.993	\\
20	&	80	&	0.2	&	0.997	&	0.997	&	0.997	&	0.999	&	0.999	&	0.999	\\
20	&	80	&	0.25	&	0.999	&	0.999	&	0.999	&	0.999	&	1.000	&	0.999	\\
\hline																	
50	&	20	&	0.025	&	0.050	&	0.391	&	0.139	&	0.102	&	0.465	&	0.238	\\
50	&	20	&	0.05	&	0.131	&	0.577	&	0.276	&	0.221	&	0.643	&	0.412	\\
50	&	20	&	0.1	&	0.388	&	0.838	&	0.600	&	0.530	&	0.877	&	0.722	\\
50	&	20	&	0.15	&	0.661	&	0.950	&	0.819	&	0.771	&	0.965	&	0.894	\\
50	&	20	&	0.2	&	0.833	&	0.983	&	0.925	&	0.903	&	0.990	&	0.963	\\
50	&	20	&	0.25	&	0.929	&	0.996	&	0.974	&	0.962	&	0.998	&	0.988	\\
\hline																	
50	&	30	&	0.025	&	0.126	&	0.270	&	0.196	&	0.211	&	0.383	&	0.318	\\
50	&	30	&	0.05	&	0.332	&	0.531	&	0.446	&	0.466	&	0.647	&	0.586	\\
50	&	30	&	0.1	&	0.753	&	0.876	&	0.830	&	0.842	&	0.925	&	0.900	\\
50	&	30	&	0.15	&	0.937	&	0.975	&	0.963	&	0.966	&	0.988	&	0.982	\\
50	&	30	&	0.2	&	0.985	&	0.995	&	0.992	&	0.993	&	0.998	&	0.996	\\
50	&	30	&	0.25	&	0.998	&	1.000	&	0.999	&	0.999	&	1.000	&	1.000	\\
\hline																	
50	&	40	&	0.025	&	0.216	&	0.327	&	0.286	&	0.334	&	0.447	&	0.415	\\
50	&	40	&	0.05	&	0.537	&	0.656	&	0.614	&	0.666	&	0.760	&	0.733	\\
50	&	40	&	0.1	&	0.917	&	0.951	&	0.942	&	0.954	&	0.974	&	0.970	\\
50	&	40	&	0.15	&	0.989	&	0.994	&	0.993	&	0.995	&	0.997	&	0.997	\\
50	&	40	&	0.2	&	0.998	&	0.999	&	0.999	&	0.999	&	1.000	&	1.000	\\
50	&	40	&	0.25	&	1.000	&	1.000	&	1.000	&	1.000	&	1.000	&	1.000	\\
\hline																	
50	&	50	&	0.025	&	0.317	&	0.402	&	0.376	&	0.443	&	0.534	&	0.512	\\
50	&	50	&	0.05	&	0.697	&	0.769	&	0.748	&	0.798	&	0.850	&	0.837	\\
50	&	50	&	0.1	&	0.974	&	0.982	&	0.979	&	0.986	&	0.992	&	0.991	\\
50	&	50	&	0.15	&	0.998	&	0.999	&	0.999	&	0.999	&	1.000	&	0.999	\\
50	&	50	&	0.2	&	1.000	&	1.000	&	1.000	&	1.000	&	1.000	&	1.000	\\
50	&	50	&	0.25	&	1.000	&	1.000	&	1.000	&	1.000	&	1.000	&	1.000	\\
\hline																	
50	&	60	&	0.025	&	0.412	&	0.480	&	0.461	&	0.539	&	0.608	&	0.592	\\
50	&	60	&	0.05	&	0.814	&	0.860	&	0.848	&	0.890	&	0.917	&	0.911	\\
50	&	60	&	0.1	&	0.993	&	0.995	&	0.995	&	0.996	&	0.997	&	0.997	\\
50	&	60	&	0.15	&	1.000	&	1.000	&	1.000	&	1.000	&	1.000	&	1.000	\\
50	&	60	&	0.2	&	1.000	&	1.000	&	1.000	&	1.000	&	1.000	&	1.000	\\
50	&	60	&	0.25	&	1.000	&	1.000	&	1.000	&	1.000	&	1.000	&	1.000	\\
\hline																	
50	&	80	&	0.025	&	0.598	&	0.645	&	0.634	&	0.720	&	0.755	&	0.747	\\
50	&	80	&	0.05	&	0.938	&	0.951	&	0.948	&	0.968	&	0.973	&	0.972	\\
50	&	80	&	0.1	&	0.998	&	0.999	&	0.999	&	0.999	&	1.000	&	1.000	\\
50	&	80	&	0.15	&	1.000	&	1.000	&	1.000	&	1.000	&	1.000	&	1.000	\\
50	&	80	&	0.2	&	1.000	&	1.000	&	1.000	&	1.000	&	1.000	&	1.000	\\
50	&	80	&	0.25	&	1.000	&	1.000	&	1.000	&	1.000	&	1.000	&	1.000	\\
\hline

\caption{\label{tab:power} \emph{Power of the standard chi-square based $Q$ test and the improved  $Q$ test for homogeneity (gamma and chi-square with $\e(Q)$ degrees of freedom approximations)  at the nominal 5\% and 10\% levels.  $I$ is the number of studies all of size $N$ equally divided between the treatment and control arms.  The effects have a mean of $\delta=0.5$ and variance of $\tau^2$.}}
\end{longtable}
\end{footnotesize}

\end{document}